\documentclass[prodmode,acmcsur]{acmsmall} 
\usepackage{url}
\usepackage{multirow}
\usepackage{array}
\usepackage{rotating}
\usepackage{amsmath}
\usepackage{graphicx}
\usepackage{caption}
\usepackage{color}
\usepackage{epstopdf}

\usepackage[ruled]{algorithm2e}

\graphicspath{ {Fig/} }

\SetAlFnt{\small}
\SetAlCapFnt{\small}
\newcommand{\nmytilde}{\raise.17ex\hbox{$\scriptstyle\mathtt{\not\sim}$}}

\begin{document}

\markboth{T. Chakraborty et al.}{Metrics for Community Analysis: A Survey}

\title{Metrics for Community Analysis: A Survey}
\author{Tanmoy Chakraborty
\affil{University of Maryland, College Park, MD, USA}
Ayushi Dalmia
\affil{International Institute of Information Technology, Hyderabad, India}
Animesh Mukherjee
\affil{Indian Institute of Technology, Kharagpur, India}
Niloy Ganguly
\affil{Indian Institute of Technology, Kharagpur, India}}


\setlength{\abovedisplayskip}{0pt}
\setlength{\belowdisplayskip}{0pt}
\setlength{\abovedisplayshortskip}{0pt}
\setlength{\belowdisplayshortskip}{0pt}

\setlength{\abovecaptionskip}{2pt}
\setlength{\belowcaptionskip}{2pt}
\setlength{\floatsep}{6pt}
\setlength{\textfloatsep}{6pt} 

\begin{abstract}
Detecting and analyzing dense groups or {\em communities} from social and information networks has attracted
immense attention over last one decade due to its enormous applicability in different domains. Community
detection is an {\em ill-defined problem}, as the nature of the communities is not known in advance. The problem
has turned out to be even complicated due to the fact that communities emerge in the network in various
forms -- disjoint, overlapping, hierarchical etc. Various heuristics have been proposed depending upon the
application in hand. All these heuristics have been materialized in the form of new {\em metrics}, which in most
cases are used as optimization functions for detecting the community structure, or provide an indication of
the goodness of detected communities during evaluation. There arises a need for an organized and detailed
survey of the metrics proposed with respect to community detection and evaluation. In this survey, we
present a comprehensive and structured overview of the start-of-the-art metrics used for the detection and
the evaluation of community structure. We also conduct experiments on synthetic and real-world networks
to present a comparative analysis of these metrics in measuring the goodness of the underlying community
structure.
\end{abstract}

\category{I.5.3}{Clustering}{Algorithms}

\terms{Design, Algorithms, Performance}

\keywords{Metrics, community discovery, community evaluation}

\maketitle

\section{Introduction}

Community structure of networks has attracted a great deal of attention of the researchers in computer science, especially in the areas of data mining and social network analysis. Although, community is an {\em ill-defined} concept \cite{Fortunato201075}, a general consensus suggests that community structure is a decomposition of nodes of  a network into sets  such that nodes within a set are densely connected internally, and sparsely connected externally \cite{Girvan2002}. Communities are formed due to the structural or functional similarities among the vertices in the network \cite{newman4}. Therefore, analyzing the  community structure of a network provides a high-level view of the formation of network structure through the interactions of nodes having identical nature. 

Communities in real-world networks are of different kinds: disjoint or non-overlapping (e.g., students belonging to different disciplines in an institute) \cite{Fortunato201075}, overlapping (e.g., person having membership in different social groups in Facebook) \cite{Xie:2013},
hierarchical (e.g., cells in human body form tissues that in turn form organs	 and so on) \cite{Balcan:2013} etc. The task of community analysis goes through two separate phases: first, {\em detection} of meaningful community structure from a network, and second, {\em evaluation} of the appropriateness of the detected community structure.  Since there is no single definition of a community which is universally accepted, many thoughts emerged which in turn resulted in different definitions of community structure. Every definition of community is  justified 
in terms of different metrics formulated. Therefore, analyzing such metrics is crucial to understand the development of  the studies pertaining to community analysis.

Due to the lack of consensus on the definition of a community, formulating metrics for evaluating the quality of a community is also a challenging task. There has been a large number of works aimed at designing functions quantifying properties proposed by their own in order to evaluate the goodness of a community. Interestingly, these quality functions are not only useful for evaluation purposes but can also be used in various methods to detect the communities. The evaluation of the communities detected by an algorithm from a network becomes easier if the actual (ground-truth) community structure of the network is known a priori. In this direction, attempts were made to construct artificially-generated networks with inherent community structure \cite{Lancichinetti}. To make a correspondence between the detected and ground-truth community structures, few metrics were borrowed from the literature of {\em clustering} in data mining \cite{p.2006survey} and reformulated by incorporating the network information. 

Although, there have been studies surveying different approaches mostly on the {\em detection} of non-overlapping \cite{PhysRevE.80.056117,Fortunato201075} and overlapping \cite{Xie:2013,citeulike:13779944} communities separately, no attempts have been initiated to understand thoroughly {\em the metrics used to design the algorithms and to evaluate the quality of the algorithms}. Metrics help us in understanding the goodness of an algorithm in a quantitative manner. We believe that evaluating a community detection algorithm may be difficult because it requires selecting between several proposed quality functions that often output contradictory results. The structural properties of the network and of the communities being looked for may strongly differ from one case to the other. It is then of utmost importance to identify the right metric depending upon the purpose and properties of the given network. Moreover, most of the metrics are extensions of some old measures. Therefore, we need to understand the evolutionary route of a derived metric and its possible extensions.

In this paper, we conduct an extensive survey on the state-of-the-art metrics used for detecting and evaluating the community structure. More importantly, we bring together metrics related to all the classes of community structures (disjoint, overlapping, fuzzy, local etc.) into a single article that can help us understand the derivation of one measure into another. Specifically, in case of metrics related to community detection we study traditional metrics like Modularity and its variants as well as more recently proposed metrics including Permanence, Surprise, Significance and Flex. In case of community evaluation, we study various state-of-the-art metrics like Normalised Mutual Information, Purity, Rand Index and F-Measure etc. Finally, a comparative analysis is presented based on the experiments conducted on synthetically-generated and real-world networks.  \\ 

\noindent{\bf Organization of the survey:} The survey has been conducted in two broad directions. In Section \ref{sec:qualityMetric}, we shall introduce metrics used as optimization functions in different algorithms for community detection in networks. In Section \ref{sec:metricsOfCommunityEvaluation}, we shall discuss the measures used for evaluating the goodness of the detected community structure (i.e., performance of the community detection algorithms). Note that these two sets of metrics (discussed in Sections \ref{sec:qualityMetric} and \ref{sec:metricsOfCommunityEvaluation}) are not mutually exclusive. In Section \ref{result},  we shall present a comparative analysis on the performance of the state-of-the-art metrics based on the experiments conducted on different networks. In each of these sections mentioned above, we shall present the metrics pertaining to all kinds of community structures. Finally in Section \ref{conc}, we conclude the survey by addressing the shortcomings and limitations of the metrics found in the literature of community detection. To facilitate the discussion, we present in Table ~\ref{table:Notations}  the summary of the notations used in this survey.

\section{Metrics for discovering community structure}
\label{sec:qualityMetric}
The goal of community detection is to find inherent communities in a network. However, the definition of a community is not clear. What is a good community? Given a graph $G(V,E)$ where $V$ is the set of nodes and $E$ is the set of edges, one can have an exponentially large number of possible communities. Enumerating these communities is an NP-Complete problem \cite{Garey1974NPComplete}. Moreover, not all partitions of a graph are equally good. In order to obtain the best partition of the graph and thus significant communities, most of the community detection algorithms aim to optimize a {\em goodness metric} which essentially indicates the quality of the communities detected from the network. The goal of the community detection algorithm would be to obtain the best partition of the network which would optimize the metric. A wide variety of such metrics have been proposed which can detect the quality of the communities obtained for a given partition. This section presents a detailed study on the large number of goodness metrics proposed by the graph 
mining algorithms for community detection. Note that apart from their extensive usage for community detection in different algorithms, these metrics are widely used  to evaluate the quality of a detected community structure. We shall discuss this issue in Section \ref{sec:metricsOfCommunityEvaluation}.
\begin{table}[!t]
\centering
\caption{Notations used in this paper.}
\scalebox{0.85}{
 \begin{tabular}{|c|c|}
 
 \multicolumn{2}{c}{{\bf Graph-specific}}\\\hline
 $G(V,E)$ & A graph with set of nodes $V$ and set of edges $E$\\
 $A$      & Adjacency matrix of graph $G$\\
 $N$       & $N=|V|$, Number of nodes in $G$\\
 $m$      &   $m=|E|$, Number of edges in $G$ \\
 $\Gamma_{u}$& Neighbors of node $u$\\
 $d(u)$   & Degree of node $u$\\
 $d(u)^{in}$   & In-degree of node $u$\\
 $d(u)^{out}$   & Out-degree of node $u$\\
 $C_{in}(u)$ & Internal clustering coefficient of node $u$\\\hline
 \multicolumn{2}{c}{{\bf Community-specific}}\\\hline
 $\Omega$ & Detected non-overlapping community structure, $\Omega=\{\omega_1,\omega_2,\cdots,\omega_K\}$\\
  $\Psi$         & Detected overlapping community structure, $\Psi=\{\psi_1,\psi_2,\cdots,\psi_K\}$\\
 $C$      & Ground-truth community structure $C=\{c_1,c_2,\cdots,c_J\}$\\
 $N_{\omega_j}$& Number of nodes in community $\omega_j$\\
 $N_{\omega_i c_j}$& $|\omega_i \cap c_j|$, Number of nodes present in both $\omega_i$ and $c_j$ communities \\
$|E_{\omega}^{in}|$ & Number of edges between the nodes within the community $\omega$\\
 $|E_{\omega}^{out}|$ & Number of edges between the nodes in community $\omega$ to nodes outside community $\omega$ \\
\hline
 \end{tabular}}
 \label{table:Notations}
\end{table}

\subsection{Metrics used for non-overlapping community detection}
 The non-overlapping community detection algorithms aim at partitioning the vertices of $G(V,E)$ into $K$ number of non-empty, mutually exclusive groups, $\Omega=\{\omega_1,\omega_2,\cdots,\omega_K\}$ such that each vertex belongs to {\em one and only one} community, i.e., $|\omega_1|+|\omega_2|+\cdots +|\omega_K|=|V|$.

Various simple metrics capturing the topological properties of the network have been proposed in the past which are used to compute the quality of communities detected. Let us consider a function $f(\omega)$ that characterizes the quality of the community on the basis of the connectivity of nodes in $\omega$. \cite{Yang:2012} summarized these scoring functions and grouped them into the following three broad classes:\\\\
\noindent (A) {\it Scoring functions based on internal connectivity}:
\vspace{-2mm}
\begin{itemize}
 \item {\bf Internal density:} $f(\omega)=\frac{|E_{\omega}^{in}|}{|\omega|(|\omega|-1)/2}$ is the internal edge density of the node in community $\omega$ \cite{radicchi2004}.
 \item {\bf Edge inside:} $f(\omega) =|E_{\omega}^{in}|$ is the number of edges between the members of community $\omega$ \cite{radicchi2004}.
 \item {\bf Average degree:} $f(\omega) = 2|E_{\omega}^{in}|/|\omega|$ is the average internal degree of the members of community $\omega$ \cite{radicchi2004}.
 \item {\bf Fraction over median degree (FOMD):} $f(\omega)=\frac{|{u:u\in \omega,|{(u,v):v\in \omega}|> d_m}|}{|\omega|}$ is the fraction of nodes of $\omega$ that have internal degree higher than $d_m$, where $d_m$ is the median value of $d(u)$ in $V$.
 \item {\bf Triangle Participation Ratio (TPR):} It is the fraction of nodes in $\omega$ that belong to a triad: $f(\omega) = \frac{|{u:u\in \omega,\{v,w\in \omega,(u,v)\in E,(u,w)\in E,(v,w)\in E\} \neq \phi}|}{|\omega|}$. 
\end{itemize}
 \noindent(B) {\it Scoring functions based on external connectivity}:
 \begin{itemize}
  \item {\bf Expansion} measures the number of edges per node that point outside the cluster: $f(\omega) = |E_{\omega}^{out}|/|\omega|$ \cite{radicchi2004}.
  \item {\bf Cut Ratio} is the fraction of existing edges (out of all possible edges) leaving the cluster: $f(\omega) =|E_{\omega}^{out}|/|\omega|(N−|\omega|)$ \cite{Fortunato201075}.
 \end{itemize}
 \noindent (C) {\it Scoring functions that combine internal and external connectivity}:
 \begin{itemize}
  \item {\bf Conductance}: $f(\omega) = \frac{|E_{\omega}^{out}|}{2|E_{\omega}^{in}|+|E_{\omega}^{out}|}$ measures the fraction of total edge volume that points outside the community \cite{Shi:2000}.
  \item {\bf Normalized Cut}: $f(\omega) = \frac{|E_{\omega}^{out}|}{2|E_{\omega}^{in}|+|E_{\omega}^{out}|} + \frac{|E_{\omega}^{out}|}{2(m - |E_{\omega}^{in}|)+|E_{\omega}^{out}|}$ normalises the cut score. \cite{Shi:2000}.
  \item {\bf Maximum-ODF (Out Degree Fraction)}: $f(\omega) = max_{u\in \omega} \frac{|{(u,v)\in E:v \notin \omega}|}{d(u)}$ is the maximum fraction of edges of a node in $\omega$ that point outside $\omega$ \cite{Flake:2000}.
  \item {\bf Average-ODF}: $f(\omega) = \frac{1}{|\omega|} \sum_{u \in \omega} \frac{|{(u,v)\in E:v \notin \omega}|}{d(u)}$ is the average fraction of edges of nodes in $\omega$ that point out of $\omega$ \cite{Flake:2000}.
\item {\bf Flake-ODF}: $f(\omega) = \frac{|{u:u\in \omega,|{(u,v)\in E:v\in \omega}|<d(u)/2}|}{|\omega|}$ is the fraction of nodes in $\omega$ that have fewer edges pointing inside than to the outside of the cluster \cite{Flake:2000}.
\end{itemize}
\noindent (D) {\it Scoring function based on a network model}:
\begin{itemize}
 \item {\bf Modularity}: It computes the difference between the fraction of edges for a given partition of the original graph and a null graph. The choice of the null graph is in principle arbitrary, and several possibilities exist. The usual choice of the null graph is to choose a model with the same degree distribution as of the original graph. For an unweighted and undirected network, modularity is defined as,
\begin{equation}
\label{eqn:modularity_actual}
Q_{ud} = \sum_{\omega \in \Omega} \Bigg[ \frac{|E_{\omega}^{in}|}{m} - \Bigg(\frac{ |E_{\omega}^{in}+ E_{\omega}^{out}|}{2m}\Bigg)^{2} \Bigg]
\end{equation}
An alternative way of defining the modularity of a graph is given as follows ~\cite{newman2006modularity}:
\begin{equation}
\label{eqn:modularity2}
Q_{ud} = \frac{1}{2 m} \sum_{ij} \Bigg[ A_{ij} - \frac{d(i)d(j)}{2m} \Bigg] \delta_{\omega_{i},\omega_{j}}
\end{equation}
where $\delta_{\omega_{i},\omega_{j}}$ is the Kronecker delta function, which returns $1$ if $\omega_{i}=\omega_{j}$, and $0$ otherwise. The value of modularity lies between -1 and 1. A higher value of modularity indicates a strong community structure. 
\end{itemize}
\cite{Yang:2012} further defined four goodness metrics $f(\omega)$ for a community $\omega$ which capture the network structure:
\begin{itemize}
 \item {\bf Separability} captures the intuition that good communities are well-separated from the rest of the network \cite{Shi:2000,Fortunato201075}, meaning that they have relatively few edges pointing from set $\omega$ to the rest of the network. Separability measures the ratio between the internal and the external number of edges of $\omega$: $f(\omega) = |E_{\omega}^{in}|/|E_{\omega}^{out}|$. 
 \item {\bf Density} is built on the intuition that good communities are well connected \cite{Fortunato201075}. It measures the fraction of the edges (out of all possible edges) that appear between the nodes in $\omega$, $f(\omega) = 2|E_{\omega}^{in}|/\omega(\omega-1)$.
 \item {\bf Cohesiveness} characterizes the internal structure of the community. Intuitively, a good community should be internally well and evenly connected, i.e., it should be relatively hard to split a community into two sub-communities. This is characterized by the conductance of the internal cut. Formally, $f(\omega) = min_{\omega^{'} \subset \omega} \phi(\omega^{'})$, where $\phi(\omega^{'})$ is the conductance of $\omega^{'}$ measured in the induced subgraph by $\omega$. Intuitively, conductance measures the ratio of the edges in $\omega^{'}$ that point outside the set and the edges inside the set $\omega^{'}$. A good community should have high cohesiveness (high internal conductance) as it should require deleting many edges before the community would be internally split into disconnected components \cite{Leskovec:2010}.
 \item {\bf Clustering coefficient} is based on the premise that network communities are manifestations of locally inhomogeneous distributions of edges, because pairs of nodes with common neighbors are more likely to be connected with each other \cite{Watts1998}.
\end{itemize}

In another paper, \cite{Leskovec:2010} used two other topological metrics to measure the quality of a community:
\vspace{-3mm}
\begin{itemize}
\item {\bf Volume}: $\sum_{u\in \omega} d(u)$ is the sum of degrees of nodes in $\omega$.
\item {\bf Edges cut}: $|E_{\omega}^{out}|$ is the number of edges needed to be removed to disconnect nodes in $\omega$ from the rest of the network. 
\end{itemize}

Among the metrics discussed above, modularity is the most widely used one to detect the strength of the communities. Introduced in the seminal paper ~\cite{newman2004finding} the principal idea behind modularity is that the number of inter-community edges for a given graph must be greater than the number of edges for a random graph having a similar degree distribution as the original graph. The definition of modularity suggested by ~\cite{newman2004finding} in Equation \ref{eqn:modularity_actual} was applicable only to \textit{unweighted} and \textit{undirected} graphs. Several modifications and extensions to modularity have been proposed in the literature of community detection. The proposed changes cater to the specific tasks and type of graphs one may intend to analyze.

\underline{\textbf{Modularity for weighted graphs:}} ~\cite{newman2004weightedModularity} proposed a simple extension to the existing definition of modularity for weighted graphs. One can view a weighted graph as a multigraph with multiple edges between a pair of nodes. Mapping the weighted graph to a multi-graph, it can be easily shown that Equation \ref{eqn:modularity2} is a generalized formula for modularity. For a weighted network, instead of $A_{ij}$, we use $W_{ij}$ representing the weight of the edge between nodes $i$ and $j$, the degree $d(i)$ is now replaced with the strength $s(i)$ of node $i$. The strength of a node is the sum of the degree of the adjacent nodes. For proper normalization, the number of edges $m=|E|$ in Equation \ref{eqn:modularity2} has to be replaced by the sum $W$ of the weights of all edges. Thus, the product $\frac{s(i)s(j)}{2W}$ is now the expected weight of the edge $(i,j)$ in the null model of modularity, which has to be compared with the actual weight $W_{ij}$ of 
that edge in the original graph. The equation of modularity for a weighted network can be written as,
\begin{equation}
\label{eqn:weightedModularity1}
Q_{w} = \frac{1}{2 |W|} \sum_{ij} \Bigg[ W_{ij} - \frac{s(i)s(j)}{2|W|} \Bigg] \delta_{\omega_{i},\omega_{j}}
\end{equation}


\underline{\textbf{Modularity for directed graphs:}} \cite{arenas2007directedModularity} and ~\cite{Leicht2008directedModularity} proposed extensions to modularity for directed graphs. ~\cite{arenas2007directedModularity} elegantly extended the definition of modularity while preserving its semantics in terms of probability to the scenario of directed networks. If an edge is directed, the probability that it will be oriented in either of the two possible directions depends on the in- and out-degrees of the end vertices. For instance, taken two vertices $i$ and $j$ under consideration, where $i$ has a high in-degree and low out-degree, while $j$ has low in-degree and high out-degree, in the null model of modularity an edge will be much more likely to point from $j$ to $i$ than from $i$ to $j$. Therefore, the expression of modularity for directed graphs can be defined as follows,
\begin{equation}
\label{eqn:directedModularity}
Q_{d} = \frac{1}{ m} \sum_{ij} \Bigg[ A_{ij} - \frac{d(i)^{out} d(j)^{in}}{m} \Bigg] \delta_{\omega_{i},\omega_{j}}
\end{equation}
where $d(i)^{out}$ and $d(i)^{in}$ are the in-degree and out-degree of node $i$ respectively. If a graph is both directed and weighted, Equations \ref{eqn:weightedModularity1} and \ref{eqn:directedModularity} can be combined as follows, which is the most general form of modularity,
\begin{equation}
\label{eqn:generalModularity}
Q_{gen} = \frac{1}{ |W|} \sum_{ij} \Bigg[ W_{ij} - \frac{s_{i}^{out}s_{j}^{in}}{|W|} \Bigg] \delta_{\omega_{i},\omega_{j}}
\end{equation}

\underline{\textbf{Similarity-based modularity:}} \cite{song2007SimilarityBased} proposed similarity-based modularity which is robust to the groups of nodes in a  graph which have dense inter-connections. Instead of using edges within communities and between communities as the criteria of partitioning they propose a more general concept, \textit{similarity} $S(i,j)$, to measure the graph partition quality. The notion of similarity between two vertices $i$ and $j$ is given by the number of shared neighbors normalized by the number of neighbors of each vertex. For a partition $\Omega$, the similarity-based modularity $(Q_{s})$ is given as,
\begin{equation}
Q_{s} = \sum_{\omega \in \Omega} \Bigg[ \frac{\sum_{i} \sum_{j} S(i,j) \delta(i,\omega) \delta_(j,\omega) }{ \sum_{i} \sum_{j} S(i,j)} - \Bigg( \frac{\sum_{i} \sum_{j} S(i,j) \delta(i,\omega)}{\sum_{i} \sum_{j} S(i,j)} \Bigg)^{2} \Bigg]
\end{equation}
where 
\[
   \delta(i,\omega) = 
\begin{cases}
    1,              & \text{if } i \in \omega\\
    0,              & \text{otherwise}
\end{cases}
\]

\underline{\textbf{Motif modularity:}} \cite{arenasMotif2008} used the underlying principle of modularity and proposed motif modularity where they use motifs instead of edges. They proposed that short paths, or motifs, of a network, could be used to define and identify both communities and more general topological classes of nodes. Communities will be defined based on the principle that they ``contain'' more motifs than a null model representing a randomized version of the network. As a particular case, the triangle modularity of a partition $\Omega$ reads,
\begin{equation}
Q^{\triangle}_{mm} (\omega) = \Bigg[ \frac{\sum_{ijk} A_{ij}(\omega) A_{jk}(\omega) A_{ki}(\omega)} {\sum_{ijk} A_{ij} A_{jk} A_{ki}} - \frac{\sum_{ijk} n_{ij}(\omega) n_{jk}(\omega) n_{ki}(\omega)} {\sum_{ijk} n_{ij} n_{jk} n_{ki}} \Bigg]
\end{equation}
where $A_{ij}(\omega) = A_{ij} \delta_{\omega_{i}, \omega_{j})}$, $n_{ij} = d(i)d(j)$, $n_{ij}(\omega) = n_{ij}\delta_{\omega_{i},\omega_{j}}$, and $\delta_{\omega_{i},\omega_{j}}=1$ if both $i$ and $j$ are part of same community, $0$ otherwise. 

\underline{\textbf{Max-Min modularity:}} \cite{ChenDetectingcommunities} presented a new measure, called max-min modularity, which considers the property of both connected $Q_{Max}$ (same as modularity) and user-defined related node pairs in finding communities $Q_{Min}$. The Max-Min modularity is given by,
\begin{equation}
\begin{split}
Q_{Max\_Min} & = Q_{Max} - Q_{Min}\\
			& = \sum_{ij} \Bigg[ \frac{1}{2m} \Bigg( A_{ij} -\frac{d(i)d(j)}{2m} \Bigg) - \frac{1}{2m^{\prime}} \Bigg( A_{ij}^{\prime} -\frac{d(i)^{\prime}d(j)^{\prime}}{2m^{\prime}}  \Bigg) \Bigg]\delta_{\omega_{i},\omega_{j}}
\end{split}
\end{equation}
The second part tries to minimize the modularity of the complement of the graph $G$, given by $G^{\prime}$, constructed taking into consideration the user-defined criteria $U$ (to define whether two disconnected nodes are related or not). The higher $Q_{Max\_Min}$ is, the better community division over $Q_{gen}$.

\underline{\textbf{Influence-based modularity:}} \cite{ghosh2010InfluenceBased} proposed influence-based modularity where they claimed that ``a community is composed of individuals who have a greater capacity to influence others within their community than outsiders.'' The influence is measured using a \textit{influence matrix} $P$ which captures the number of $n$-$hop$ length paths between nodes $i$ and $j$ for all the pairs of nodes in the graph. If $\bar{P}$ indicates the expected capacity to influence, then influence based modularity is given by,
\begin{equation}
Q_{i} = \sum_{ij} \Bigg[ P_{ij} - \bar{P_{ij}}\Bigg] \delta_{\omega_{i}, \omega_{j}}
\end{equation}

\underline{\textbf{Diffusion-based modularity:}} \cite{Kim2010modifiedModularity} remarked that the directed modularity of Equation \ref{eqn:directedModularity} may not properly account for the directedness of the edges, and proposed a modified definition of modularity based on diffusion on directed graphs, inspired by Google's PageRank algorithm. They proposed the concept of \textit{LinkRank} which indicates the importance of links instead of nodes as in the case of PageRank algorithm. LinkRank is the probability that a random walker is moving from node $i$ to $j$ in the stationary state. By using LinkRank, the modified definition of modularity can be written as,
\begin{equation}
\label{eqn:linkRankModularity}
Q_{lr} =  \sum_{ij} \Bigg[ L_{ij} - E(L_{ij}) \Bigg] \delta_{\omega_{i},\omega_{j}}
\end{equation}
where $E(L_{ij})$ is the expected value of $L_{ij}$ in the null model. In Equation \ref{eqn:linkRankModularity}, it is easy to notice that the first term is the fraction
of time spent on walking within communities by a random walker since $L_{ij}$ is the probability of the random walker following the link from $i$ to $j$, and the second term is the expected value of this fraction in a null model.

\underline{\textbf{Dist-modularity:}} \cite{liu2012extending} extended the existing definition of modularity and proposed Dist-modularity. Dist-modularity captures the similarity attraction feature in the null model. In the new null model the expected number of edges is given by,
\begin{equation}
P_{ij}^{Dist} = \frac{\tilde{P_{ij}}+\tilde{P_{ji}}}{2} 
\end{equation}
where,
\begin{equation}
\tilde{P_{ij}} = \frac{d(i) d(j) e^{-(s_{ij}/\sigma)^2}}{\sum_{v \in V} d(v)e^{-(s_{iv}/\sigma)^2} } 
\end{equation}
where $s_{ij}$ denotes the similarity distance between $v_{i}$ and $v_{j}$ -- the smaller the $s_{ij}$, the more similar are the two nodes. $\sigma$ is a parameter of the quality metric that controls how fast or slow the function $e^{-(s_{ij}/\sigma)^2}$ decreases. 
Using the new null model, dist-modularity is given by,
\begin{equation}
Q_{Dist} = \frac{1}{2 m} \sum_{ij} \Bigg[ A_{ij} - P_{ij}^{Dist} \Bigg] \delta_{\omega_{i},\omega_{j}}
\end{equation}

\underline{\textbf{Limits of modularity:}} Given a network, the knowledge of the size of the communities is not known a priori. Modularity and modularity-based definitions discussed so far are not robust in nature and would fail to capture communities of all types in a network. \cite{fortunato02007Limits} discussed the limits of modularity. They found that modularity optimization may fail to identify modules smaller than a scale which depends on the total size of the network and on the degree of interconnectedness of the modules, even in cases where modules are unambiguously defined. This is known as the {\em resolution limit} of modularity. ~\cite{Good2010Performance} studied modularity at a deeper level and pointed out the degeneracy problem of modularity. Modularity may not have a global minima. There are typically an exponential number of structurally diverse alternative partitions with modularities very close to the optimum, often known as the {\em degeneracy problem}. This problem is most severe when applied to networks with modular structure; it occurs for weighted, directed, bipartite and multi-scale generalizations of modularity; and it is likely to exist in many of the less popular partition score functions for module 
identification. However, several  modifications of modularity have been proposed in the past to address the limits of modularity. In the following section we briefly discuss the modifications proposed which overcome the limits of the original definition.

\underline{\textbf{Methods to overcome the limitations of modularity:}}  Modifications of modularity's null model were introduced by \cite{massen2005identifying}  and ~\cite{muff2005localmodularity}. ~\cite{massen2005identifying} identified the sensitivity of modularity towards larger communities. The value of modularity tends to decrease with the increase in the large number of small communities as the expected number of edges is larger than the actual edges in the graph. This leads to a bias toward detecting large communities. In order to address this, they proposed a modified form of modularity by improving the predicted fraction of edges within communities. The method is to create an ensemble of random networks with the same degree distribution as the original and with the constraint that multiple edges and self-connections are forbidden. They first calculated the probability of an edge between any pair of nodes in a such an ensemble network. The predicted fraction of edges within each community in a random network is given by the sum of these probabilities over each pair of nodes within the community. This sum (denoted by $f_{\omega}$) replaces the second element of modularity $Q$ in Equation \ref{eqn:modularity_actual}. Thus, {\bf modified modularity} is now given by,
\begin{equation}
Q_{rnl} = \sum_{\omega \in \Omega} \Bigg[ \frac{|E^{in}_{\omega}|}{m} - f_{\omega} \Bigg]
\end{equation}

~\cite{muff2005localmodularity} addressed the limits of modularity by proposing a local version of modularity, in which the expected number of edges within a module is not calculated with respect to the full graph, but considering just a portion of it, namely the subgraph including the module and its neighboring modules. Their motivation is the fact that modularity's null model implicitly assumes that each vertex could be attached to any other, whereas in real cases a community is usually connected to few other communities. On a directed graph, their \textbf{localized modularity}
$LQ$ reads,
\begin{equation}
LQ = \sum_{\omega \in \Omega} \bigg[ \frac{|E^{in}_{\omega}|}{L_{{\omega}n}} - \frac{|E^{in}_{\omega}| |E^{out}_{\omega}|}{L_{{\omega}n}^2} \bigg]
\end{equation}
where $L_{{\omega}n}$ is the total number of edges in the subgraph comprising community $\omega$ and its neighbor communities. The localized modularity is not bounded by $1$, but can take any value. 

In order to address the resolution limit ~\cite{statistical2006Reichardt} used parameters to tune the contribution of the null model. They scaled the topology by a factor $r$ by adding self loops of the same magnitude $r$ to the vertices.

~\cite{Li2008Quantitative} proposed a quantitative measure for evaluating the partition of a network into communities based on the concept of average modularity degree called the \textbf{modularity density} or \textbf{D-value}. For a partition $\Omega$ the modularity density is given by,
\begin{equation}
D = \sum_{\omega \in \Omega} d(G_{\omega})
\end{equation}
where $d(G_{\omega})$ is the average modularity degree of subgraph $G_{\omega}=(V_{\omega}, E_{\omega})$ given by,
\begin{equation}
 d(G_{\omega}) =  d_{in}(G_{\omega}) -  d_{out}(G_{\omega})
\end{equation}
where $d_{in}(G_{\omega})$ and  $d_{out}(G_{\omega})$ are the average inner and outer degrees of subgraph $G_{\omega}$ respectively.

~\cite{arenas2008Resolution} viewed the resolution limit of modularity as a feature of the quality function and proposed a modification to the modularity function that allows to view the network at multiple resolutions. The modularity of the network at scale $r$ is given by,
\begin{equation}
Q_{r} = \sum_{\omega \in \Omega} \Bigg[ \frac{|E_{\omega}^{in}|+|V_{\omega}|r}{|E|+|V|r} - \Bigg(\frac{ |E_{\omega}^{in}+ E_{\omega}^{out}|+|V_{\omega}|r}{2|E|+|V|r}\Bigg)^{2} \Bigg]
\end{equation}
The resolution limit of the method is supposedly solvable with the introduction of modified versions of the measure, with tunable resolution parameters. However, ~\cite{lancichinetti2011Limits} showed that multi-resolution modularity suffers from two opposite coexisting problems: the tendency to merge small subgraphs, which dominates when the resolution is low; the tendency to split large subgraphs, which dominates when the resolution is high.

~\cite{ye2012newDefinition} argued that simply counting the number of edges within a module is not enough to capture the topology of a network. In order to capture the topology of the network they proposed a new parameter, called minimal diameter $D_{\omega}$ which is defined as the average minimal path for all pairs of vertices in a given module. 
The new definition of modularity metric is then given as:
\begin{equation}
Q_{d} = \sum_{\omega \in \Omega} \Bigg[ \frac{|E_{\omega}^{in}|}{D_{\omega}} - \Bigg(\frac{ |E_{\omega}^{in}+ E_{\omega}^{out}|}{2|E|}\Bigg)^{2}*\frac{1}{\tilde{D_{\omega}}} \Bigg]
\end{equation}
where $D_{\omega}$ is the observed diameter for the community $\omega$ and can be computed easily. $\tilde{D_{\omega}}$ is the expected diameter for a randomised graph and is computed by the methods discussed by ~\cite{fronczak2004averagePath}.

~\cite{Sun2013ModularityIntensity} addressed the resolution limit of modularity by proposing the quality metric \textbf{modularity intensity}. Maximizing modularity intensity can resolve the resolution limit problem, and the model effectively captures the community evolutionary process. It evaluates the cohesiveness of a community, which not only considers links between vertices, but also link weights. A community with a higher modularity intensity indicates that it is hard to split or die out. 

~\cite{chen2013ModularityDensity}  addressed the multi-resolution problem of modularity where in some cases it tends to favor small communities over large ones while in others, large communities over small ones. They first proposed modularity with \textbf{split penalty} given by,
\begin{equation}
\label{eqn:modifiedModularity}
Q_{s}=Q_{ud}-SP
\end{equation}
The split penalty addresses the problem of favoring small communities by measuring the negative effect of edges joining members of different communities and is given by,
\begin{equation}
\label{eqn:splitPenalty}
SP = \sum_{\omega_{i} \in \Omega} \Bigg[ \sum_{\substack{\omega_{j} \in \Omega \\ \omega_{j} \neq \omega_{i}}} \frac{|E_{\omega_{i},\omega_{j}}|}{2m}\Bigg]
\end{equation}
where  $|E_{\omega_{i},\omega_{j}}|$ is the number of edges (sum of weights of edges) from community $\omega_{i}$ to community $\omega_{j}$ for unweighted (weighted) networks. We can use Equation \ref{eqn:modularity_actual} and Equation \ref{eqn:splitPenalty} in Equation \ref{eqn:modifiedModularity} to obtain the modified modularity with split penalty. Using split penalty alone may lead to large communities which may not be desirable. In order to address this they introduced {\bf community density}. For undirected networks, the proposed metric \textbf{modularity density} is given by,
\begin{equation}
\begin{split}
Q_{uds} = \sum_{\omega_{i} \in \Omega} \Bigg[ \frac{|E_{\omega_{i}}^{in}|}{m}d_{\omega_{i}} - \Bigg(\frac{ |E_{\omega_{i}}^{in}+ E_{\omega_{i}}^{out}|}{2m}d_{\omega_{i}}\Bigg)^{2} - \sum_{\substack{\omega_{j} \in \Omega \\ \omega_{j} \neq \omega_{i}}} \frac{|E_{\omega_{i},\omega_{j}}|}{2m}d_{\omega_{i},\omega_{j}}\Bigg]
\end{split}
\end{equation}
where $d_{\omega_{i}}$ is the internal density of community $\omega_{i}$ and $d_{\omega_{i},\omega_{j}}$ is the pair-wise density between communities $\omega_{i}$ and $\omega_{j}$.

\cite{Zhang2012CommunityIdentification} proposed a new community metric given as follows,
\begin{equation}
\Phi(\Omega) = \Phi_{1}(\Omega) - \Phi_{2}(\Omega)
\end{equation}
where,
\begin{equation}
\Phi_{1}(\Omega) = \sum_{\omega \in \Omega}\frac{|E_{\omega}^{in}+E_{\omega}^{out}|}{N_{\omega}}, \Phi_{2}(\Omega) = \sum_{\omega_{i} \in \Omega} \sum{i \neq j} \frac{|E_{\omega_{i},\omega_{j}|}}{N_{\omega_{i},\omega_{j}}}
\end{equation}
The function $\Phi_{1}(\Omega)$ defines the sum of the average degrees in each subnetwork and $\Phi_{2}(\Omega)$defines the sum of the average number of connections between one subnetwork and other subnetworks. It is easy to see that for community identification, our goal is to both maximize $\Phi_{1}(\Omega)$ and minimize $\Phi_{2}(\Omega)$.

\cite{miyauchiZModularity2015} addressed the resolution limit of modularity by proposing a new quality metric called \textbf{Z-Modularity}. For  a division $\Omega$ they quantified the statistical rarity of division $\Omega$ in terms of the fraction of the number of edges within communities. To this end, they considered the following edge generation process over $V$. First $N$ edges are placed over $V$ at random with the same distribution of vertex degree. Then, the probability that the edge is placed within communities is given by,
\begin{equation}
p=\sum_{\omega \in \Omega} {\Bigg( \frac{D_{\omega}}{2m} \Bigg)}^{2}
\end{equation}
where $D_{\omega}$ is the sum of degrees of all the nodes in $\omega$. 
Note that this edge generation process is the same as the null-model used in the definition of modularity, with the exception of the sample size. Unlike the null-model, the sample size $N$ is not necessarily equal to the number of edges $m$. Let $X$ be a random variable denoting the number of edges generated by the process within communities. Then, X follows the binomial distribution $B(N, p)$. By the central limit theorem, when the sample size $N$ is sufficiently large, the distribution of $X/N$ can be approximated by the normal distribution $N (p, p(1-p)/N)$. Thus, we can quantify the statistical rarity of division $\Omega$ in terms of the fraction of the number of edges within communities using the Z-score as follows,
\begin{equation}
\label{eqn:zscore}
Z{(\Omega)}= \frac{ \sum_{\omega \in \Omega} \frac{|m_{\omega}| }{m}  - \sum_{\omega \in \Omega} {\Bigg( \frac{|D_{\omega}|}{2m} \Bigg)}^{2}  }{\sqrt{ \sum_{\omega \in \Omega} {\Bigg( \frac{|D_{\omega}|}{2m}\Bigg)}^{2} {\Bigg( 1-  \sum_{\omega \in \Omega} {\Bigg( \frac{|D_{\omega}|}{2m}\Bigg)}^{2} \Bigg)}^{2} }}
\end{equation}
The sample size $N$ never depends on a given division; thus, it is omitted in the denominator. 

Recently, ~\cite{xiang2015multi} proposed multi-resolution methods by using the generalized self-loop rescaling strategy to address the resolution limit of modularity.

\underline{\textbf{Adaptive scale modularity:}} \cite{Twan2014Axioms} proposed six properties (see Section 1 of SI Text for details) as axioms for community-based quality functions viz. permutation invariance, scale invariance, richness, monotonicity, locality and continuity. They analyzed these six properties and found that modularity is permutation invariant, scale invariant and continuous. In order to adapt modularity to obey the remaining two properties, they proposed adaptive scale modularity. Instead of normalizing the edge-weights with the sum of the edge weights, they proposed to normalize it using a constant $M$. However, if one only keeps the factor $M$, the fixed scale modularity does not obey monotonicity. The parameter $\gamma \geq 2$ ensures that the adaptive scale modularity obeys all six axioms. The proposed metric is give by,
\begin{equation}
Q_{asm} = \sum_{\omega \in \Omega} \Bigg[ \frac{|E_{\omega}^{in}|}{M+ \gamma (|E_{\omega}^{in}+ E_{\omega}^{out}|)} - \Bigg(\frac{ |E_{\omega}^{in}+ E_{\omega}^{out}|}{M+ \gamma (|E_{\omega}^{in}+ E_{\omega}^{out}|)}\Bigg)^{2} \Bigg]
\end{equation} 


Until now, we have discussed modularity and various adaptations of the metric to incorporate different properties of the network and the community. The metrics discussed so far follow the underlying principle of modularity; for a good community, the number of edges within the community should be larger than the number of edges in a random graph which follows the same degree distribution as the original graph. We now pay attention to other community detection metrics.
\\
\underline{{\bf Community Score:}} \cite{Pizzuti2008CD} proposed a simple but effective goodness function that maximizes the in-degree of the nodes belonging to the community and that implicitly minimizes their out-degree. Consider a community $\omega$ where $\mu_{i}$ denote the fraction of edges connecting node $i$ to the other nodes in $\omega$. The power mean of $\omega$ of order $r$, denoted as $M(\omega)$ is defined as,
\begin{equation}
M(\omega) = \frac{\sum_{i \in \omega}(\mu_{i})^{r}}{N_{\omega}}
\end{equation}
The volume $v(\omega)$ of a community $\omega$ is given by $|E_{\omega}|$.
 The score of a community $\omega$ is given by $score(\omega) = M(\omega)v_{\omega}$. The community score of the graph corresponding to a partition $\Omega$ is given by,
\begin{equation}
CS = \sum_{\omega \in \Omega} score(\omega)
\label{communityscore}
\end{equation}

\underline{{\bf SPart:}} \cite{chira2012Evolutionary} proposed a new fitness function, SPart. The quality of a network partition in communities is evaluated by looking at the contribution of each node and its neighbors to the strength of its community via the internal and external degrees. Given a node $v$ belonging to a community $\omega$, a node score evaluating the contribution of the node to the strength of its community is computed using the following measure,
\begin{equation}
SNode(v) = \frac{d(v)^{in} - d(v)^{out}}{|\omega|}
\end{equation}
A high positive value of $SNode(v)$ indicates an important contribution of node to the strength of its community. The fitness or goodness of a community $\omega$ is evaluated by considering the contribution $SNode(x)$ of each node $x \in \omega$ (first level nodes) and furthermore the contribution of all nodes $w \in \omega$ (second level nodes) which are linked with $v$. The measure used to compute the fitness of each community is defined as follows,
\begin{equation}
SComm(\omega) = \sum_{v \in \omega} \Bigg [ SNode(v) + \frac{1}{2} \sum_{\substack{w \in \omega \\ A_{vw}=1} } SNode(w) \Bigg]
\end{equation}
The overall fitness of a particular partition $\Omega$ is given as follows,
\begin{equation}
SPart(\Omega) = \frac{1}{|\Omega|} \sum_{i \in \Omega} SComm(\omega_{i}) \frac{ \nu(\omega_{i})}{|\omega_{i}|}
\end{equation}
where $\nu(\omega_{i})$ is the ratio of the number of edges within the community $\omega$ to the total number of edges.

\underline{{\bf Significance:}} \cite{traag2013significant} highlighted the fact that we do not want to know the probability a `fixed' partition containing at least $E$ internal edges, but whether a partition with at least $E$ internal edges can be found in a random graph. The probability for finding a certain partition can be reduced to finding some dense subgraphs in a random graph. The probability that a subgraph $S$ of size $n_{\omega}$ and density $q$ appears in a random graph $\mathcal{G}$ of size $n$ and density $p$ is asymptotically,
\begin{equation}
	Pr(S(n_{\omega},q) \subseteq \mathcal{G}(n,p))= e^{\Theta (- {{n_{\omega}\choose{k}} D(q||p) })}
\end{equation}
where $D(q||p)$ is the Kullback-Leibler divergence.
For a partition $\Omega$ we can compute the probability for the partition to be contained in a random graph as follows:
\begin{equation}
Pr(\Omega) = \prod_{\omega} exp (- {{n_{\omega}}\choose{2}} D (p_{c}||p))
\end{equation}
Significance is then given by,
\begin{equation}
S(\Omega) = - \log Pr(\Omega) = \sum_{c} {{n_\omega} \choose 2} D(p_\omega || p)
\end{equation}

\underline{{\bf Permanence:}} \cite{chakraborty2014permanence} proposed permanence, a vertex based community quality metric to quantify the propensity of a vertex to remain in its assigned community and the extent to which it is ``pulled'' by the neighboring communities. The permanence of a vertex $v$ compares the internal connection of a vertex $I(v)$ with the maximum connections to a single external community, i.e.,  $E_{max}(v)$. This is normalized by the degree of the vertex $d(v)$. Alongside, permanence also captures how well the vertex is connected within the community via internal clustering coefficient $c_{in}(v)$.  This
criterion  emphasizes that a vertex is likely to be within a community if it is a part of a near-clique substructure. Mathematically, the permanence of a vertex $v$ is given by,
\begin{equation}
Perm(v) = \Bigg[\frac{I(v)}{E_{max}(v)}\times \frac{1}{d(v)}\Bigg] - \Bigg[ 1 - c_{in}(v) \Bigg]
\end{equation}
The sum of the permanence of all vertices, normalized by the number of vertices, provides the permanence of the network. It indicates to what extent, on an average, the vertices of a network are bound to their communities. The value of permanence ranges from $+1$ to $-1$. Permanence has been proved to ameliorate  the limitations of modularity.

\underline{{\bf Surprise:}} Another community quality metric, Surprise proposed by ~\cite{aldecoa2011Surprise} compares the distribution of the nodes and links in communities in a given network with respect to a null model. Surprise assumes as a null model that links between nodes emerge randomly. It then evaluates the departure of the observed partition from the expected distribution of nodes and links into communities given the null model via KL-divergence. To do so, it uses the following cumulative hypergeometric distribution,
\begin{equation}
S= \sum_{j=p}^{Min(M,|E|)} \frac{ {{M} \choose{j}}{{F-M}\choose{|E|-j}}}{{{F}\choose{|E|}}}
\end{equation}
where $F$ is the maximum possible number of links in a network, $E$ is the observed set of links, $M$ is the maximum possible number of intra-community links for a given partition, and $p$ is the total number of intra-community links observed in that partition. Using a cumulative hypergeometric distribution allows to calculate the exact probability of the distribution of links and nodes in the communities defined for the network by a given partition. Thus, Surprise measures how unlikely (``surprising'') is that distribution. Unlike modularity, Surprise also considers the role of number of units within each community along with the number of links. Qualitatively, Surprise performed better than modularity. However, this claim was later confirmed by ~\cite{aldecoa2013surprise,aldecoa2014tool,Fleck2014Surprise}. In practice, it is not straightforward to work with, nor is it simple to implement in an optimization procedure, mainly due to numerical computational problems.

~\cite{Traag2015asymptoticSurprise} proposed an asymptotic approximation of Surprise by allowing the links to be withdrawn with replacement. Thus, \textbf{asymptotical Surprise} is given by,
\begin{equation}
S= \sum_{j=p}^{Min(M,|E|)} {{M} \choose{j}} {\langle q \rangle }^{j} (1-{\langle q \rangle})^{M-j}
\end{equation}
where $q=\frac{\sum_{\omega \in \Omega}|E_{\omega}^{in}|}{|E|}$ and ${\langle q \rangle }$ indicates the expected value of $q$. 

\underline{{\bf Expected nodes:}} \cite{Mangioni2015expected} proposed a novel quality metric, called expected nodes where they studied the link partitions as oppose to the node partitions in modularity. Intuitively, a group of links $\omega$ is a relevant community if it consists of a large number of links adjacent to a few nodes $(V_{in})$. A node is an internal node of $\omega$ if one of its stubs (half-links) is in $\omega $. Therefore, to compute the expected number of internal nodes in the configuration model, one can choose randomly $2|\omega|$ stubs among a total of $2|E|$ stubs. A node $u$ has therefore $b(u)$ ways to be picked.The expected number of internal nodes for a given link group $\omega$, denoted by $\mu_{G}(|\omega|)$, is then,
\begin{equation}
\mu_{G}(|\omega|) = \sum_{u \in V} 1 - \frac{{{2|E|-b(u)} \choose {2|\omega|}}}{{{2|E|} \choose {2|\omega|}}}
\end{equation}
 A group $\omega$ has a good internal quality if it has less internal nodes than expected. The internal quality $Q_{in}$ for a given group $\omega$  is given as the variation between the actual number of internal nodes $V_{in}(\omega)$ and its expectation $\mu_{G}(|\omega|)$,
\begin{equation}
Q_{in}(\omega) = \frac{\mu_{G}(|\omega|)-V_{in}(\omega)}{\mu_{G}(|\omega|)}
\end{equation}
Using similar arguments, one can write the external quality of a group $\omega$ as,
\begin{equation}
Q_{ext}(\omega) = \min \Bigg( 0, \frac{|V_{out}(\omega)| - \mu_{G \setminus \omega} (\bar{d}(\omega)/2)  }{\mu_{G \setminus \omega} (\bar{d}(\omega)/2)} \Bigg)
\end{equation}
where $\bar{d}(\omega)=\sum_{u \in V_{in}(\omega)}\bar{d}(\omega,u)$; $\bar{d}(\omega,u)$ be the degree of $u$ restricted to links not in $\omega$. 
Thus, the expected nodes for a group $\omega$ is defined as,
\begin{equation}
Q_{\omega}= 2 \frac{|\omega|Q_{in}(\omega)+|\omega_{out}|Q_{ext}(\omega)}{|\omega|+|\omega_{out}|}
\end{equation}
The formulation for expected nodes for a given partition $\Omega$ is given as the weighted sum of the quality of each group,
\begin{equation}
Q_{G}(\Omega) = \frac{\sum_{\omega \in \Omega} |\omega| Q(\omega) }{|E|}
\end{equation}

\underline{{\bf Communitude:}} \cite{Miyauchi2015NetworkCommunity} quantified the community degree of $\omega$ in terms of the fraction of the number of edges within the subgraph induced by $\omega$. Inspired by the metric, Z-score, proposed in Equation~\ref{eqn:zscore}, a similar probability distribution of the fraction of the number of edges within the subgraph generated using the nodes in $\omega$ is estimated. Next, they quantified the community degree of $\omega$ in terms of the fraction of the number of edges within the subgraph using the Z-score as follows,
\begin{equation}
com(\omega)=\frac{\frac{|E_{\omega}^{in}|}{m} - (\frac{|E_{\omega}^{in}+E_{\omega}^{out}|}{2m})^2}{(\frac{|E_{\omega}^{in}+E_{\omega}^{out}|}{2m})^2(1-(\frac{|E_{\omega}^{in}+E_{\omega}^{out}|}{2m})^2)}
\end{equation}
If $\omega=\phi$, $com(\omega)=0$. The upper bound for the metric is $1$. Note, this quality function can be viewed as a modified version of the function given by Z-score, which is normalized by the standard deviation of the fraction of the number of edges within the subgraph.

\cite{CreusefondLP14} used {\bf compactness} which measures the potential speed of a diffusion process in a community. Starting from the most eccentric node, the function captures the number of edges reached per time step by a perfect transmission of information. The underlying model defines community as a group of people within which communication quickly reaches everyone. They recently used it as a community-level quality function to measure the goodness of the detected community \cite{CreusefondLP15}.

The {\bf local internal clustering coefficient} \cite{Watts1998} is also used as a quality metric in \cite{CreusefondLP15}. It is defined by the probability that two neighbors of a vertex that are in the same community are also neighbors. The clustering property of communities is actually one of the most well-known in this field, and is explained by the construction of social networks by {\em homophily} \cite{mcpherson2001birds}. 

A summary of the metrics discussed in this subsection can be found in the SI Text (Table I).

 \vspace{-5mm}

 \subsection{Metrics for overlapping community detection}
In real world networks, it is well-understood that a node can be naturally characterized by multiple community memberships. For instance, a person in social network can have connections to several groups, such as friends, family and colleagues; a researcher may be active in several areas. Moreover, in social networks the number of communities a vertex can belong to is unlimited because there is no restriction on the number of communities a person can be a part of. This phenomenon also happens in other networks, such as biological networks \cite{thai2011handbook} where a node might have multiple functions. Therefore, overlapping community structure is a natural phenomenon in real networks, and thus there has been an increasing interest to discover communities  that are not necessarily disjoint.

More formally, given a graph $G(V,E)$, the task of overlapping community detection is to partition the vertices $V$ into non-empty sets $\Psi = \{\psi_{1} , \psi_{2} , ..., \psi_{|\Psi|} \}$, where a vertex can belong to more than one set, i.e., $|\psi_{1}|+ |\psi_{2}|+\cdots+|\psi_{|\Psi|}| \geq |V|$.

Note that in this case, the relationship between a node and a community is binary, i.e., a node either {\em completely} belongs to a community or does not, resulting \textit{crisp overlapping community}. While on the other hand, several attempts have been conducted with the assumption that each node is associated with communities in proportion to a belongingness factor. This is called \textit{fuzzy overlapping community} structure. In this section, we discuss the metrics associated with both these notions of overlapping community detection. 

Although initial attempt for overlapping community detection was done by ~\cite{palla2005uncovering},
\cite{zhang2007identification} introduced the original notion of overlapping community structure. Given a set of overlapping communities $\Psi = \{\psi_{1} , \psi_{2} , ..., \psi_{|\Psi|} \}$ in which a node may belong to more than one of them, a vector of belonging coefficients $(\alpha_{i\psi_{1}} , \alpha_{i\psi_{2}} , ..., \alpha_{i\psi_{|\Psi|}} )$ can be assigned to each node $i$ in the network. A belonging coefficient $\alpha_{i\psi}$ measures the strength of association between node $i$ and community $\psi$. Without loss of generality, the following constraints are assumed to hold,
\begin{equation}
\label{eqn:belongingnessVector}
0 \leq \alpha_{i\psi } \leq 1  \quad  \forall i \in V, \forall \psi \in \Psi \qquad and \qquad \sum_{\psi \in \Psi} \alpha_{i\psi} = 1
\end{equation}

\if{0}
However, overlaps can be of two types: crisp and fuzzy. With non-fuzzy or crisp overlapping, each vertex belongs to one or more communities with equal strength: a node either belongs to a community or it  does not. With fuzzy overlapping, each node may also belong to more than one community but the strength of its membership to each community can vary. Here, we discuss the metrics associated with both overlapping and fuzzy communities.
\fi

\underline{{\bf Modularity:}} Using the notion of fuzzy overlap ~\cite{zhang2007identification} defined the membership of each community as $ \bar{V_{\psi}} = \{i| \alpha_{i\psi} > \lambda, i \in V \} $, where $\lambda$ is a threshold that can convert a soft assignment into final clustering. They proposed a generalized notion of modularity as follows,
\begin{equation}\label{q_z}
Q_{ov}^{z} = \sum_{\psi \in \Psi} \Bigg[ \frac{|E_{\psi}^{in}|}{m} - \Bigg(\frac{2|E_{\psi}^{in}|+|E_{\psi}^{out}| }{2m}\Bigg)^{2} \Bigg]
\end{equation}
where $|E_{\psi}^{in}| = \sum_{i,j \in \psi} ((\alpha_{i\psi}+\alpha_{j\psi})/2) W_{ij}$ and $|E_{\psi}^{out}| = \sum_{i \in \psi j \in N-N_{\psi}} (\alpha_{i\psi}+(1-\alpha_{j\psi})/2) W_{ij}$. This modified modularity was used to give a relative membership of the nodes.

~\cite{nepusz2008fuzzy} extended modularity for computing the goodness of an overlapping partition $\Psi$ by replacing the Kronecker delta function in Equation \ref{eqn:modularity2} and proposed a fuzzy variant of modularity as follows, 
\begin{equation}
Q_{ov}^{F}= \frac{1}{2m} \sum_{ij} \Bigg[ A_{ij} - \frac{d(i)d(j)}{2m} \Bigg] s_{ij}
\end{equation}
where,
\begin{equation}
s_{ij} = \sum_{\psi \in \Psi} \alpha_{i\psi}\alpha_{j\psi}
\end{equation}
The goal is thus to find a fuzzy partition such that the difference between the actual similarity among the vertices given by the adjacency matrix and the fuzzy similarity given by the belongingness vector is minimized.

A similar goodness metric was proposed by ~\cite{shen2009overlap} where they extended modularity as follows:
\begin{equation}
\label{eqn:shenModularity}
Q_{ov}^{S} =  \frac{1}{2m}  \sum_{ij} \Bigg[ A_{ij} - \frac{d(i)d(j)}{2m} \Bigg] \alpha_{i\psi} \alpha_{j\psi}
\end{equation}

~\cite{Shen20091706} proposed to optimize Equation \ref{ee}, to detect both the overlapping and hierarchical properties of complex community structure together by introducing a belonging coefficient. The belonging coefficient $\alpha_{i\psi}$ of a node $i$ for a given community is redefined as the number of communities $O_{i}$ to which it belongs. The extended modularity of the overlapping community structure is given by,
\begin{equation}\label{ee}
Q_{ov}^{E} =  \frac{1}{2 m} \sum_{\psi \in \Psi} \sum_{ij} \Bigg[ A_{ij} - \frac{d(i)d(j)}{2m} \Bigg] \frac{1}{O_{i}O_{j}}
\end{equation}

~\cite{nicosia2009Extension} proposed the following measure expressed in terms of a function $F$,
\begin{equation}\label{q_n}
Q_{ov}^{N} = \frac{1}{2m} \sum_{\psi \in \Psi} \sum_{ij} \Bigg[ A_{ij} F \Big( \alpha_{i\psi},\alpha_{j\psi} \Big) - \frac{d(i)d(j) \Bigg( \sum_{v \in V} F \Big(\alpha_{v\psi}, \alpha_{j\psi} \Big) \Bigg) \Bigg( \sum_{v \in V} F \Big(\alpha_{i\psi}, \alpha_{v\psi}\Big) \Bigg)}  {2mN^{2}} \Bigg] 
\end{equation}
where $F(\alpha_{i\psi},\alpha_{j\psi})$ could be defined as a product $\alpha_{i\psi}\alpha_{j\psi}$, an average $(\alpha_{i\psi}+\alpha_{j\psi})/2$, a maximum $max(\alpha_{i\psi},\alpha_{j\psi})$, or any other suitable function.

~\cite{Lancichinetti2009OverLapping} maximized the following local fitness function in their proposed algorithm LFM to obtain natural communities,
\begin{equation}
f(\psi) = \frac{|E_{in}|^{\psi}}{(|E_{in}|^{\psi}+|E_{out}|^{\psi})^\alpha}
\label{fitness}
\end{equation}
where $\alpha$ is  the resolution parameter controlling the size of the communities. 

~\cite{Pizzuti2009OCD} used the definition of community score (Equation ~\ref{communityscore}) introduced in ~\cite{Pizzuti2008CD} on the line graph of the given network $G$.

~\cite{lazar2010modularity} defined a crisp overlapping goodness measure as follows for a partition $\Psi$:
\begin{equation}
Q_{ov}^{crisp} = \frac{1}{|\Psi|}\sum_{\psi \in \Psi} Q_{\psi}
\end{equation}
The modularity $Q_{\psi}$ for a given community $\psi$ is given by,
\begin{equation}
Q_{\psi} = \frac{|E_{\psi}^{in}|+|E_{\psi}^{out}|}{N_{\psi}(N_{\psi}-1)/2} \frac{1}{N_{\psi}} \sum_{i \in N_{\psi}} \frac{ \sum_{j \in N_{\psi} i \neq j} A_{ij} - \sum_{j \notin N_{\psi}} A_{ij} } {d(i)s_{i}}
\end{equation}
Here $s_{i}$ is the number of communities in which node $i$ belongs to.

~\cite{chen2010Overlapping} proposed using the modified modularity $Q_{ov}^{\Psi}$ for weighted networks defined as,
\begin{equation}
Q_{ov}^{\Psi} =  \frac{1}{2m} \sum_{\psi \in \Psi} \sum_{ij} \Bigg[ A_{ij} - \frac{d(i)d(j)}{2m} \Bigg] \alpha_{i\psi} \alpha_{j\psi}
\end{equation}
where $\alpha_{i\psi} = \frac{k_{i\psi}}{\sum_{\psi \in \Psi}k_{i\psi}}$ is the strength with which node $i$ belongs to community $\psi$, and $k_{i\psi} = \sum_{j \in \psi} w_{ij}$ is the total weight of links from $i$ into community $\psi$.

~\cite{havemann2011Overlapping} proposed the following modification to the fitness function given by Equation~\ref{fitness},
\begin{equation}
f(\psi) = \frac{|E_{in}|^{\psi}+1}{(|E_{in}|^{\psi}+|E_{out}|^{\psi})^\alpha}
\end{equation}
which allows a single node to be considered a community by itself. This avoids violation of the principle of locality.

~\cite{chen2014modularityDensity} extended the existing definition of modularity density introduced in ~\cite{chen2013ModularityDensity}. They proposed the following goodness function,
\begin{equation} \label{q_md}
\begin{split}
Q_{ov}^{MD} & = \sum_{\psi \in \Psi} \Bigg[ \frac{|E_{\psi}^{in}|}{m}d_{\psi} - {\Bigg( \frac{ 2|E_{\psi}^{in}|+|E_{\psi}^{out}| }{2m} d_{\psi} \Bigg)}^{2} - \sum_{\psi^{'} \in \Psi, \psi \neq \psi^{'}} \frac{|E_{\psi,\psi^{'}}|} {2m} d_{\psi,\psi^{'}} \Bigg],  \\
d_{\psi} & =  \frac{2|E_{\psi}^{in}|}{\sum_{i,j \in \psi, i \neq j} f(\alpha_{i\psi},\alpha_{j\psi})}, \\
d_{\psi, \psi^{'}} & =  \frac{|E_{\psi, \psi^{'}}|}{\sum_{i \in \psi,j \in \psi^{'}} f(\alpha_{i\psi},\alpha_{j\psi})}
\end{split}
\end{equation}
where $|E_{\psi}^{in}| = \frac{1}{2} \sum_{i,j \in \psi} f(\alpha_{i\psi},\alpha_{j\psi})A_{ij},  |E_{\psi}^{out}| = \sum_{i \in \psi}\sum_{ \substack{\psi^{'} \in \Psi \\ \psi \neq \psi^{'} \\ j \in \psi^{'}}} f(\alpha_{i\psi},\alpha_{j\psi^{'}})A_{ij},  $ and $  |E_{\psi,\psi^{'}}| = \sum_{i \in \psi,j \in \psi^{'}} f(\alpha_{i\psi},\alpha_{j\psi^{'}})A_{ij} $.

\underline{{\bf Flex:}} \cite{mangioni2015Flex} introduced a novel quality metric, called flex which tries to balance two objectives at the same time: maximize both the number of links inside a community and the local clustering coefficient of each community. The first step to calculate flex for a given partition of the network is to
define the \textit{Local Contribution} of a node $i$ to a given community $\psi$,
\begin{equation}
LC(i,\psi) = \lambda* \triangle(i,\psi) + (1-\lambda)*N(i,\psi) - \kappa * \wedge(i,\psi),
\end{equation}
where $\triangle(i,\psi)$ is the ratio between the transitivity of node $i$ (number of triangles that $i$ forms) inside community $\psi$ and the total transitivity of this node in the full network, $N(i,\psi)$ is the ratio between the number of neighbors node $i$ has inside community $\psi$ and its total number of neighbors, and $\wedge(i,\psi)$ is the ratio between the number of open triangles in community $\psi$ that contain node $i$ and the total participation of $i$ in the whole network. Variables $\lambda$ and $\kappa$ are weights that balance the importance of each term. 

Given the local contribution of all nodes to each community, the \textit{Community Contribution} (CC) of a community $\psi$ in a given partition is defined as,
\begin{equation}
CC(\psi) = \sum_{i \in \psi} LC(i,\psi) - \frac{{N_{\psi}}^{\gamma}}{N},
\end{equation}
where $\gamma$ is the penalization weight devised to avoid the generation of a trivial solution in which the entire network forms a single community. The \textit{Flex} value of a given partition $\Psi$ is given by,
\begin{equation}\label{flex}
Flex(\Psi) = \frac{1}{N} \sum_{\psi \in \Psi} CC(\psi)
\end{equation}

~\cite{baumes2005finding} proposed a two-step process which maximizes the following local density function,

\begin{equation}
f(\psi) = \frac{|W_{in}|^{\psi}}{|W_{in}|^{\psi}+|W_{out}|^{\psi}}
\end{equation}

A modified version was introduced by \cite{kelley2009existence} after incorporating the edge probability $e_{p}$, where the parameter $\lambda$ controls how the algorithm behaves in sparse areas of the network.

\begin{equation}
f(\psi) = \frac{|W_{in}|^{\psi}}{|W_{in}|^{\psi}+|W_{out}|^{\psi}} + \lambda e_{p} 
\end{equation}

A summary of the metrics discussed in this subsection can be found in the SI Text (Table II).

\subsection{Other metrics for community detection}

The majority of the community quality detection metrics which have been proposed in the literature pertain to overlapping and non-overlapping communities only. However, attempts have been made to propose community quality metrics depending upon the application in hand or the network properties. Table \ref{tab:otherMetrics} summarizes the different metrics proposed so far. A detailed explanation of all the metrics mentioned in Table \ref{tab:otherMetrics} can be found in the SI text (Section 2).

\begin{table}[!h]
\centering
\tbl{Other metrics for community detection.\label{tab:otherMetrics}}
{
\begin{tabular}{|l|l|}
\hline
Task            & Metrics \\
\hline\hline
Local Community Detection & Local Modularity (\cite{clauset2005finding}) \\
                          & Subgraph Modularity (\cite{Luo:2006}) \\
                          & Density Isolation (\cite{Lang:2007}) \\
                          & Community Internal Relation (\cite{5231879})\\
                          & Community External Relation (\cite{5231879})\\
                          & Local-Global Fitness Function (\cite{Lancichinetti})\\
                          & Internal Density (\cite{SahaHKRZ10})\\
                          & Conductance (\cite{Andersen:2006,Kloster:2014})\\
                          & Edge-Surplus (\cite{Tsourakakis:2013})\\
\hline\hline
$n$-partitie networks & Bipartite Modularity (\cite{Barber2007modularity,Guimer2007modularity}) \\
                          & Modified Bipartite Modularity (\cite{Murata2009Bipartite}) \\
                          & Tripartite Modularity (\cite{neubauer2009towards}) \\
                          & Tripartite Modularity (\cite{murata2010detecting,murata2011tripartite}) \\
                          & Mutli-faceted Bipartite Modularity (\cite{Suzuki2009MultiFacet}) \\
                          & Density-Based Bipartite Modularity (\cite{Xu2015278}) \\
                          & Intensity Score (\cite{Max-Intensity}) \\
\hline\hline
Multiplex Networks & Modularity for Multiplex Networks (\cite{tang2009multilayer}) \\
                          & Redundancy (\cite{berlingerio2011multiDimensional}) \\
                          & Cross-Layer Edge Clustering Coefficient (\cite{aberer2012multiLayer}) \\
\hline\hline
Signed Networks & Map Equation for Signed Networks (\cite{esmailian2015community})     \\                     
\hline\hline
Anti-Community Detection & Anti-Modularity (\cite{Chen2014293}) \\
\hline                      
\end{tabular}}
\end{table} 

\section{Metrics for Community Evaluation}\label{val}
\label{sec:metricsOfCommunityEvaluation}

Once the communities from a network are detected using a community detection algorithm, the next task is to evaluate the detected community structure. The evaluation becomes easier, if the actual community structure of the network is given to us (we call the actual community structure of a network as ``ground-truth'' community structure). In such cases, various metrics can be used to make a correspondence between the detected and the ground-truth communities. We refer to these metrics as ground-truth based validation metrics. If the algorithm is able to detect a community structure which has high resemblance with the ground-truth, the algorithm is well-accepted and can be used further for other networks where the underlying ground-truth communities might not be available. 

\subsection{Ground-truth based validation metrics for non-overlapping community structure}\label{ground_truth_metric}
In this section,  we discuss the metrics used to measure the similarity between the detected and the ground-truth community structures for non-overlapping community structure. Note that  most of these metrics are borrowed from the literature of ``clustering'' in data mining.

Let us recall the notations to be used in this section again: given a graph $G(V,E)$,  $\Omega=\{\omega_1,\omega_2,\cdots,\omega_K\}$ is the set of detected communities, and $C=\{c_1,c_2,\cdots,c_J\}$ is the set of ground-truth communities. $N=|V|=\sum_{k\in K} |\omega_k|=\sum_{j\in J} |c_j|$ is the total number of nodes, $N_{c_j}=|c_j|$ and $N_{\omega_i c_j}=|\omega_i \cap c_j|$. 

\cite{Lin2006,Manning:2008} introduced {\bf purity} where each detected community is assigned to the ground-truth label which is most frequent in the community. Formally, it is defined as,
\begin{equation}
Purity (\Omega,C)=\frac{1}{N}\sum_k \underset{j}{\max} |\omega_k,c_j|
\end{equation}
The upper bound is $1$; it corresponds to a perfect match between the partitions. The lower bound is $0$ and indicates the opposite.
Note that purity is not a symmetric measure: processing the purity of $\Omega$ relatively to $C$ amounts to considering the parts of $\Omega$ majority in each part of $C$. Therefore, in general, there is no reason to consider that $Purity (\Omega,C)$ and $Purity (C,\Omega)$ are equal. In cluster analysis, the former version is generally used, and called simply {\em purity}, whereas the second version is the {\em inverse purity} \cite{Artiles:2007}. \cite{Danon} made a comment, explaining how it can be biased by the number and sizes of communities. However, their remark is actually valid only for the purity, and not for the inverse version.
  
High purity is easy to achieve when the number of communities is large; in particular, purity is $1$ if each node gets its own community. In contrast, the inverse purity favors algorithms detecting few large communities. The most extreme case occurs when the algorithm puts all the nodes in the same community. Thus, we cannot use purity to trade off the quality of the community against the number of communities. To solve this problem, Newman introduced an additional constraint \cite{PhysRevE.69.066133}: when an estimated community is majority in several actual communities, all the concerned nodes are considered as misclassified.  The solution generally adopted in cluster analysis rather consists in processing the {\bf F-Measure}, which is the harmonic mean of both the versions of the purity \cite{Artiles:2007}.
  \begin{equation}\label{f-score}
   F-Measure=\frac{2\cdot Purity(\Omega,C) \cdot Purity(C,\Omega)}{Purity(\Omega,C) + Purity(C,\Omega)}
  \end{equation}

Another interpretation of community is to view it as a series of decisions, one for each of the   pairs of nodes in the network \cite{hubert1985}.  Two nodes will be assigned to the same community if and only if they both have same label in ground-truth. A true positive ($TP$) decision assigns two same-labeled nodes to the same community; a true negative ($TN$) decision assigns two different labeled nodes to different communities. There are two types of errors we can commit. A ($FP$) decision assigns two different labeled nodes to the same community. A ($FN$) decision assigns two same labeled nodes to different communities. The {\bf Rand index} ($RI$) measures the percentage of decisions that are correct and is given by,
\begin{equation}
RI=\frac{TP+TN}{TP+FP+FN+TN}
\end{equation}

RI gives equal weight to FPs and FNs. We can use the $F_{\beta}$ to penalize FNs more strongly than FPs by selecting a value $\beta>1$, thus giving more weight to recall (R) as follows,
\begin{equation}
P=\frac{TP}{TP+FP};\ R=\frac{TP}{TP+FN};\ F_{\beta}=\frac{(\beta^2+1)PR}{\beta^2P+R} 
\end{equation}

Although RI is more stringent and reliable, it has some shortcomings. In brief, RI can be shown to have biases that may make its results misleading in certain application scenarios. There is a family of other external indexes that can be used in order to get more accurate results, such as the Jaccard coefficient \cite{Halkidi:2001}, the Minkowski measure \cite{Jiang:2004}, the Fowlkes–Mallows index \cite{Fowlkes}, and the $\tau$ statistics \cite{Jain:1988}. 

In the domain of community detection, the chance-corrected version of RI, called {\bf Adjusted Rand Index} (ARI) \cite{hubert1985}, is often preferred. It seems to be less sensitive to the number of communities \cite{Vinh:2009}. The chance correction is based on the general formula defined for any measure $M$,
\begin{equation}\label{m}
 M_c=\frac{M-E(M)}{M_{max}-E(M)}
\end{equation}
where $M_c$ is the chance-corrected measure, $M_{max}$ is the maximal value that $M$ can reach, and $E(M)$ is the value expected for some null model. According to \cite{hubert1985}, under this assumption that the partitions are generated randomly with the constraint of having fixed number of communities and part sizes, the expected value for the number of pairs in a community intersection $\omega_i \cap c_j$ is given by,
\begin{equation}
 E\dbinom{N_{\omega_ic_j}}{2}=\dbinom{N_{\omega_i}}{2}\dbinom{N_{c_j}}{2}/\dbinom{N}{2}
\end{equation}
By replacing in Equation \ref{m} and after some simplifications, we get the final ARI
\begin{equation}\label{ari}
 ARI(\Omega,C)=\frac{\sum_{ij}\dbinom{N_{\omega_ic_j}}{2} - \sum_i\dbinom{N_{\omega_i}}{2} \sum_j \dbinom{N_{c_j}}{2} / \dbinom{N}{2}}{\frac{1}{2} \bigg(\sum_i\dbinom{N_{\omega_i}}{2} + \sum_j \dbinom{N_{c_j}}{2} \bigg) - \sum_i\dbinom{N_{\omega_i}}{2} \sum_j \dbinom{N_{c_j}}{2} / \dbinom{N}{2}}
\end{equation}

Like RI, this metric is symmetric. Its upper bound is $1$, meaning both partitions are exactly similar. Because it is chance-corrected, a value equal or below $0$ represents  the fact that the similarity between $\Omega$ and $C$ is equal or less than what is expected from two random partitions.

{\bf Normalized Mutual Information} (NMI) \cite{Manning:2008,Strehl:2003,FortunatoL09}, another alternative information-theoretic metric is defined as follows:
\begin{equation}\label{nmi}
NMI(\Omega,C)=\frac{I(\sigma,C)}{[H(\sigma)+H(C)]/2}
\end{equation}
where $I$ is mutual information,
\begin{equation}\label{i}
I(\Omega,C)=\sum_k\sum_j \frac{|\omega_k \cap c_j|}{N}\ log\frac{{N|\omega_k \cap c_j|}}{|\omega_k||c_j|}
\end{equation}

$H$ is entropy as defined below,
\begin{equation}\label{h}
H(\Omega)=-\sum_k \frac{|\omega_k|}{N}\ log\frac{\omega_k}{N}
\end{equation}

NMI is always a number between 0 and 1. A major problem of NMI is that it is not a {\em true metric}, i.e., it does not follow triangle-inequality (see Section 3.2 of SI Text). 

In contrast, {\bf Variation of Information} (VI) \cite{Meila:2007,vi} or shared information distance obeys the triangle inequality. It is defined as
\begin{equation}
\begin{split}
 VI(\Omega,C)&=-\sum_{i,j}r_{ij}[ \log (r_{ij}/p_i) + \log (r_{ij}/q_j)] \\
             & =H(\Omega)+H(C)-2I(\Omega,C)
\end{split}              
\end{equation}
where $p_i=|\omega_i|/N$, $q_j=|c_j|/N$ and $r_{ij}=|\omega_j \cap c_j|/N$. A perfect agreement with a known structure will provide a value of $VI$ = 0.

However, \cite{Orman12} argued (see Section 3.1 of SI Text) that the traditional metrics consider a community structure as simply a partition, and  therefore ignore a part of the available information: the network topology.  In order to make a more reliable evaluation, \cite{Orman12} proposed to jointly use traditional metrics and various topological properties. However, they also acknowledged that this makes the evaluation process more complicated, due to the multiplicity of values to take into account. Recently, \cite{abs-1303-5441}  proposed the solution which consists in retaining a single value, by modifying traditional measures so that they take the network topology into account. The proposed metrics are: modified purity, modified ARI and modified NMI.

To begin with, a notion of purity of {\em a node} is defined for a partition $\Omega$ relatively to another partition $C$:
\begin{equation}
 Purity(u,\Omega,C)=\delta(C_j|\underset{~\forall j} {\mathrm{s.t.}} ~N_{\omega_i c_j}\ is\ maximum)
\end{equation}
where $u\in \omega_i$ and $u\in c_j$; and $\delta$ is the Kronecker delta function. The function is therefore binary: $1$ if the community of $C$ containing $u$ is majority in that of $\Omega$ also containing $u$, and otherwise. The purity of a part $\omega_i$ relatively to a partition $C$ can then be calculated by averaging the purity of its nodes: $Purity(\omega_i,C)=\frac{1}{|\omega_i|} \sum_{u\in \omega_i} Purity (u,\Omega,C)$. So, for all the partitions in $\Omega$ relative to $C$, we get $Purity(\Omega,C)=\sum_i \sum_{u\in \omega_i} \frac{1}{N} Purity (u,\Omega,C)$. Here, one can notice the purity of each node is weighted by a value $N$.  In order to take into account the topological information, \cite{abs-1303-5441}  proposed to replace this uniform weight by a value $w_u$. Its role is to penalize more strongly misclassification concerning topologically important nodes. Therefore, the {\bf modified purity} can be defined as follows:
\begin{equation}
 Purity_M(\Omega,C)=\sum_i \sum_{u\in \omega_i} \frac{w_u}{w} Purity (u,\Omega,C)
\end{equation}
where $W=\sum_v w_v$, i.e., sum of all the weights of the nodes. This normalization allows keeping the measure between $0$ and $1$. Similarly, using the modified definition of purity, we can obtain modified F-measure using Equation \ref{f-score}. 

However, since Rand Index is based on pairwise comparisons, it is not possible to isolate the individual effect of each node, as we have seen above. Therefore, \cite{abs-1303-5441} used similarly for pairs of nodes, and proposed to use the product of the two corresponding nodal weights: $w_uw_v$. Then for any subset of nodes $S$, it can be translated as: $W(S)=\sum_{u,v\in S} w_uw_v$. Therefore, from Equation \ref{ari} the {\bf modified ARI} can be obtained as follows: 
\begin{equation}
 ARI_M(\Omega,C)=\frac{\sum_{ij}W(\omega_i\cap W(c_j) - \sum_j W(\omega_i)W(c_j)/W(V)} {\frac{1}{2} \big( \sum_i W(\omega_i) + \sum_j W(c_j) \big) - \sum_i W(\omega_i) \sum_j W(c_j) / W(s) }  
\end{equation}

Similarly, the assumption that  all nodes have the same probability $1/⁄N$ to be selected in the NMI, \cite{abs-1303-5441}  is
replaced it by the node-specific weight $w_u$. We can consequently define a modified joint probability distribution $p_{ij}^{'}=\sum_{u\in \omega_i \cap c_j} w_u/W$. Therefore, Equations \ref{i} and \ref{h} can be replaced as follows:
\begin{equation}\label{ii}
I(\Omega,C)=\sum_k\sum_j \frac{W(\omega_k \cap c_j)}{W}\ log\frac{{W\cdot W(\omega_k \cap c_j)}}{W(\omega_k)W(c_j)}
\end{equation}
\begin{equation}\label{hh}
H(\Omega)=-\sum_k \frac{W(\omega_k)}{W}\ log\frac{W(\omega_k)}{W}
\end{equation}
By replacing the above two equations in Equation \ref{nmi}, one can get the {\bf modified NMI}. 

All the modified metrics discussed above require the definition of an individual weight $w_u$ for node $u$. \cite{abs-1303-5441} considered three types of weights: (i) degree measure, $w_u=d_u/\underset{~v} max~ (d_v)$, where $d_u$ is the degree of $u$, (ii) embeddedness measure \cite{pone.0011976}, $w_u=e_u/d_u$, where $e_u$ is the internal degree of $u$ in its own community, and (iii) weighted embeddedness measure, $w_u=e_u/\underset{~v} max ~ (d_v)$. 

\cite{5520221} proposed {\bf edit distance} between pair of communities to measure the similarity. This distance counts the number of transformations needed to move from a partition $A$ to a partition $B$. This distance has two advantages and one disadvantage: it gives the matching and it is more intelligible but its matching is a one to one association. There are no merge, split or appearance of new communities, and thus the association is only reliable when such one to one association exists, i.e., only when the communities are really stable (see Section 3.3 in SI Text).

A summary of the metrics discussed in this subsection can be found in the SI Text (Table III).

\subsection{Ground-truth based validation measures for overlapping community structure}
Let us again assume that for a network $G(V,E)$,  $\Psi=\{\psi_1,\psi_2,\cdots,\psi_K\}$ is the set of detected communities, and $C=\{c_1,c_2,\cdots,c_J\}$ is the set of ground-truth communities. $N=|V|=|\cup_{k\in K} \psi_k|=|\cup_{j\in J} c_j|$ is the total number of nodes, $N_{c_j}=|c_j|$ and $N_{\omega_i c_j}=|\omega_i \cap c_j|$.

NMI was further extended for overlapping community structure \cite{Lancichinetti,McDaid}. For each node $i$ in the detected community structure $\Psi$, its community membership can be expressed as a binary vector of length $|\Psi|$, where $(x_i)_k$ is $1$ if node $i$ belongs to the $k^{th}$ cluster $\psi_k$, otherwise $0$. The $k^{th}$ entry of this vector can be viewed as a random variable $X_k$, whose probability distribution is given by $P(X_k=k)=N_k/N$, $P(X_k=0) = 1- P(X_k=1)$, where $N_k=|\psi|$, and $N$ is the number of nodes in the graph. The same holds for the random variable $Y_l$ associated with the $l^{th}$ cluster in community structure $C$. Both the empirical marginal probability distribution $P(X_k)$ and the joint probability distribution $P(X_k,Y_l)$ are used to further define entropy $H(X)$ and $H(X_k,Y_l)$. The conditional entropy of a cluster $X_k$ given $Y_l$ is defined as $H(X_k|Y_l)=H(X_k,Y_l)-H(Y_l)$. The entropy of $X_k$ with respect to the entire vector $Y$ is based on the 
best matching between $X_k$ and any component of $Y$ given by
\begin{equation}
 H(X_k|Y)=min_{l \in 1,2,...,|C|}~H(X_k|Y_l)
\end{equation}
The normalized conditional entropy of a community $X$ with respect to $Y$ is
\begin{equation}
 H(X|Y)=\frac{1}{C} \sum_k \frac{H(X_k|y)}{H(X_k)}
\end{equation}
Similarly, we can define $H(Y|X)$. Finally the NMI for overlapping community, {\bf Overlapping Normalized Mutual Information} (ONMI) for two community structures $\Omega$ and $C$ is given by $ONMI(X|Y) = 1 - [H(X|Y)+ H(Y|X)]/2$. ONMI can be easily reduced to NMI when there is no overlap in the network.

The overlapping version of the Adjusted Rand Index is {\bf Omega index}  \cite{collins1988,Murray:2012}. It is based on pairs of nodes in agreement in two community structures. Here, a pair of nodes is considered to be in agreement if they are clustered in exactly the same number of communities (possibly none). That is, the Omega index considers how many pairs of nodes belong together in no communities, how many are placed together in exactly one community, how many are placed in exactly two communities, and so on. Omega index is defined in the following way \cite{Gregory2011,havemann2011Overlapping},
\begin{equation}
 Omega(\Psi,C)=\frac{Omega_u(\Psi,C)-Omega_e(\Psi,C)}{1-Omega_e(\Psi,C)}
\end{equation}
The unadjusted Omega index $Omega_u$ is defined as,
\begin{equation}
 Omega_u(\Psi,C)=\frac{1}{M}\sum_{j=1}{max(|\Psi|,|C|)} |t_j(\psi_i) \cap t_j(c_j)|
\end{equation}
where $M=N(N-1)/2$, i.e., all possible edges, $t_j(C)$ is the set of pairs that appear exactly $j$ times in a community $C$. The expected Omega index in the null model $Omega_e$ is given by,
\begin{equation}
 Omega_e(\Psi,C)=\frac{1}{M^2}\sum_{j=1}{max(|\Psi|,|C|)} |t_j(\psi_i)|\cdot|t_j(c_j)|
\end{equation}
The larger the Omega index, the better the matching between two community structures. A value of $1$ indicates perfect matching. When there is no overlap, the Omega index reduces to the ARI.

 \cite{Campello2007833} extended the concept of RI by making it able to evaluate an overlapping partition of a data set. Further, \cite{Campello2010966} proposed another effective solution, named {\bf Generalized external index} (GEI) to the problem of comparing two overlapping partitions  by deriving first the concepts of agreements and disagreements for each individual pair of nodes ($i$,$j$). To do so, the following auxiliary definitions are needed:
 \begin{itemize}
  \item  $\alpha_{\Psi}(i,j)$: Number of communities shared by nodes $i$ and $j$ in partition $\Psi$
  \item $\alpha_{C}(i,j)$: Number of communities shared by nodes $i$ and $j$ in partition $C$.
  \item  $\beta_{\Psi}(i)$: Number of communities to which node $i$ belongs in $\Psi$, minus 1.
  \item  $\beta_{C}(i)$: Number of communities to which node $i$ belongs in $C$, minus 1.
    \end{itemize}
   
 Based on the definitions above, the agreements $a_G$ and disagreements $d_G$ associated to pair $(i,j)$ are defined as: 
 
\begin{equation}
 a_G(i,j)=min\{\alpha_{\Psi}(i,j),\alpha_{C}(i,j)\} +  min\{\beta_{\Psi}(i),\beta_{C}(i)\} + min\{\beta_{\Psi}(j),\beta_{C}(j)\}
 \end{equation}
 \begin{equation}
 d_G(i,j)=abs[\alpha_{\Psi}(i,j)-\alpha_{C}(i,j)] +  abs[\beta_{\Psi}(i)-\beta_{C}(i)] + abs[\beta_{\Psi}(j)-\beta_{C}(j)]\\
\end{equation}

These measures, in their turn, can be used to define the generalized external index  for comparing overlapping partitions: 
 \begin{equation}
  GEI(\Psi,C)=\frac{a_G}{a_G+d_G}
 \end{equation}

 \cite{HullermeierR09} proposed another extension of RI, named as {\bf Fuzzy Rand Index}. They considered RI as a distance measure. Given a fuzzy partition $P = \{P_1,P_2,...P_k\}$ of $V$, each element $v \in V$ can be characterized by its membership vector $P(v)=\{P_1(v),P_2(v),...,P_k(v)\}\in [0,1]^{k}$, where $P_i(v)$ is the degree of membership of $v$ in the $i^{th}$ community $P_i$. A fuzzy equivalence relation on $V$ can be defined in terms of a similarity measure on the associated membership vectors: $E_p(u,v)=1-||P(u)-P^{'}(v)||$. Now, given two fuzzy partitions $\Psi$ and $C$, the idea is to generalize the concept of concordance as follows. We consider a pair $(u,v)$ as being concordant in so far as $\Psi$ and $C$ agree on their degree of equivalence. This suggests to define the degree of concordance as $1-|E_{\Psi}(u,v)-E_C(u,v)|\in [0,1]$. Analogously, the degree of discordance is $|E_{\Psi}(u,v)-E_C(u,v)|$. Therefore, the distance measure on fuzzy partitions is then defined by the 
normalized sum of degrees of discordance:
 \begin{equation}
  d(\Psi,C)=\frac{\sum_{u,v\in V}|E_{\Psi}(u,v)-E_C(u,v)|}{n(n-1)/2}
 \end{equation}
Likewise, $1-d(\Psi,C)$ corresponds to the normalized degree of concordance and, therefore, is another generalization of the original Rand index.

\cite{Yang:2013} used average {\bf F1-score} to measure the equivalence of two overlapping partitions. It is defined to be the average of the F1-score of the best-matching ground-truth community to each detected community, and the F1-score of the best-matching detected community to each ground-truth community:
\begin{equation}
 F1=\frac{1}{2} (\frac{1}{|\Psi|}  \sum_{\psi_i \in \Psi} F1(\psi_i,C_{g(i)}) + \frac{1}{|C|}  \sum_{c_i \in C} F1(\Psi_{g^{'}(i)},c_i ))
\end{equation}
where the best matching $g$ and $g^{'}$is defined as follows: $g(i)=\underset{j} {\mathrm{argmax}}~F1(\Omega_i,C_j)$, $g^{'}(i)=\underset{j} {\mathrm{argmax}}~F1(\psi_j,C_i)$.

\cite{Yang:2013} also used {\bf accuracy in the number of communities} to be the relative accuracy between the detected and the true number of communities as follows: $1-\frac{|\Psi|-|C|}{2|C|}$.

Precision, Recall and F-measure (see Equation \ref{f-score}) are also used to compare two overlapping partitions \cite{Whang:2013}. \cite{Wang} used {\bf sensitivity}, {\bf specificity} and {\bf accuracy} for community evaluation. Sensitivity relates to the ability to identify the actual overlapping nodes, and is given by ratio of actual overlapping nodes to the detected overlapping nodes. Specificity relates to the ability to identify non-overlapping nodes, and is given by the ratio of actual non-overlapping nodes to all detected non-overlapping nodes. Accuracy is a ``balanced accuracy'', which is the sum of sensitivity and specificity with equal importance.

Note that for all the metrics discussed above, higher values mean more ``accurately'' detected communities, i.e., the detected node community memberships better correspond to ground-truth node community memberships. Maximum value of $1$ is obtained when the detected communities perfectly correspond to the ground-truth communities.

A summary of the metrics discussed in this subsection can be found in the SI Text (Table IV).

\section{Experiments and results}\label{result}
In this section, we demonstrate the effectiveness of the state-of-the-art community scoring functions as an indicator to measure the goodness of the community structure. In particular, we concentrate on the metrics used for evaluating non-overlapping and overlapping community structures. First, we discuss the benchmark datasets used in this experiment. Following this, we elaborate the experimental setup and the results for non-overlapping and overlapping community structures.

\subsection{Benchmark datasets}
We take both the synthetic and the real-world networks whose ground-truth community structure is known a priori.

\subsubsection{Datasets with non-overlapping community structure}
We examine a set of artificially generated networks and three real-world complex networks used in \cite{chakraborty2014permanence}.\\

\noindent {\underline{{\bf Synthetic networks:}}} We select the LFR benchmark model \cite{Lancichinetti} to generate artificial networks with a community structure.  The model allows to control directly the following properties: number of nodes $n$, desired average degree $k$ and maximal degree $k_{max}$, exponent $\gamma$ for the degree distribution, exponent $\beta$ for the community size distribution, and mixing coefficient $\mu$. The parameter $\mu$ represents the desired average proportion of links between a node and the nodes located outside
its community, called \emph{inter-community links}.  In this experiment, we vary the number of nodes ($n$) and mixing coefficient ($\mu$) to get different network structures.  For the rest of the
parameters, we use the default value of the parameters mentioned in the 
implementation\footnote{\url{https://sites.google.com/site/santofortunato/inthepress2}} designed by \cite{Lancichinetti}. Note that for each parameter configuration,
we generate 100 LFR networks, and the values in all the experiments are reported by averaging the results. \\

\noindent {\underline{{\bf Real-world networks:}}} We use three real-world networks whose properties are summarized in Table~\ref{tab:dataset}. 

\textbf{Football network}, proposed by \cite{GN} contains the network of American football games between Division IA colleges during the 
regular season of Fall 2000. The vertices in the graph represent teams (identified by their college names) and edges represent regular-season games between the two teams they connect. The teams are divided into conferences (indicating communities) containing around 8-12 teams each. Games are more frequent  between members of the same conference than between members of different conferences. Inter-conference play is not uniformly distributed; teams that are geographically close to one another but belong to different conferences are more likely to play one another than teams separated by large geographic distances.

\textbf{Railway network}, proposed by \cite{Ghosh} consists of nodes representing railway stations in India, where two stations $s_i$ and $s_j$ are connected by an edge if there exists at least one train-route such that both $s_i$ and $s_j$ are scheduled halts on that route.  Here the communities are states/provinces of India since the number of trains within each state is much higher than the trains in-between
two states. 

\textbf{Coauthorship network} is derived from the citation dataset\footnote{\url{http://cnerg.org/}} proposed by \cite{ChakrabortySTGM13,0002SGM14}. Here each node represents an author and an undirected edge between authors is drawn if the two authors collaborate at least once via publishing a paper. The communities are marked by the research fields since authors have a tendency to collaborate with other authors within the same field.

\begin{table}[!h]
\centering
\tbl{Properties of real-world networks. $n$ and $e$ are the number of nodes and edges, $c$ is the number of communities, $<k>$ and
$k_{max}$ its average and maximum degree, $n_c^{min}$ and $n_c^{max}$ the sizes of its smallest and largest
communities.\label{tab:dataset}}{
\begin{tabular}{l||r|r|r|r|r|r|r}
\hline
Network            & $n$ & $e$ & $<k>$ & $k_{max}$ & $c$ & $n_c^{max}$ & $n_c^{min}$\\
\hline\hline
Football           & 115 & 613 & 10.57 & 12 & 12 & 13 & 5  \\
Railway           & 301 & 1,224 & 6.36 & 48 & 21 & 46 & 1 \\
Coauthorship     & 103,677 & 352,183 & 5.53 & 1,230 & 24 & 14,404 & 34 \\
\hline
\end{tabular}}
\end{table}

\subsubsection{Datasets with overlapping community structure}\label{dataset}
We also examine various synthetic and real-world networks whose ground-truth communities are overlapping in nature.\\

\noindent {\underline{{\bf Synthetic networks:}}}  We use the same LFR benchmark networks proposed by \cite{Lancichinetti}. Along with the other parameters mentioned earlier, we can control two other parameters, namely the percentage of overlapping nodes $O_n$, and the number of communities to which a node belongs $O_m$. We vary the following parameters depending upon the experimental need: $n$, $\mu$, $O_n$ and $O_m$. \\

\noindent {\underline{{\bf Real-world networks:}}} We use three real networks with known overlapping ground-truth community structures\footnote{The datasets are taken from \url{http://snap.stanford.edu/}} \cite{Yang:2012}. The properties of these networks are
summarized in Table \ref{dataset_over}. 

{\bf LiveJournal} network contains nodes which are users in the LiveJournal\footnote{\url{http://www.livejournal.com/}} blogging community and edges are friendship relationships. This site also allows users form a group which other members can then join. These user-defined groups are considered as ground-truth communities.

{\bf Amazon} network is based on ``Customers Who Bought This Item Also Bought'' feature of the Amazon website\footnote{\url{www.amazon.com}}. If a product $i$ is frequently co-purchased with product $j$, the graph contains an undirected edge from $i$ to $j$. Each product category provided by Amazon defines each ground-truth community.

{\bf Youtube} network contains nodes corresponding to the users in Youtube\footnote{\url{https://www.youtube.com}}, and edges are formed due to the friendship with each other. Users can create groups which other users can join and these groups form the ground-truth communities.

\begin{table*}
\caption{Properties of the real-world networks. $n$: number of nodes, $e$: number of edges, $C$: number of
communities, $\rho$: average edge-density per community, $S$: average
size of a community, $\bar O_m$: average number of community memberships per node.}\label{dataset_over}
\centering
\scalebox{0.8}{
\begin{tabular}{l||r|r|l|r|r|r}
\hline
Network & $n$ & $e$ & C & $\rho$ & S & $\bar O_m$ \\\hline\hline
LiveJournal &  3,997,962 & 34,681,189 & 310,092 & 0.536  & 40.02 & 3.09 \\
Amazon & 334,863 & 925,872   & 151,037 & 0.769 & 99.86  & 14.83 \\
Youtube & 1,134,890 &	2,987,624 & 8,385 & 0.732 & 43.88 & 2.27 \\\hline
\end{tabular}}
\end{table*}

\subsection{Experimental setup}\label{setup}
Here we discuss the experiment conducted to evaluate the quality of the scoring metrics discussed in Section \ref{sec:qualityMetric}. Since the validation metrics discussed in Section \ref{val} compare the detected community structure directly with the ground-truth results, this evaluation is considered to be more perfect and thus preferred widely. However, as mentioned earlier for most of the real-world networks, the underlying ground-truth community structure is unknown. Therefore, the scoring metrics are used for the evaluation. Here, we intend to show to what extent a particular scoring function is able to reproduce the results obtained from the validation metrics.

In particular, we  use the framework discussed in \cite{Steinhaeuser:2010}. Let us assume that $SM=\{SM_i\}$ and $VM=\{VM_j\}$ are sets of scoring and validation metrics respectively. $CD=\{CD_1,CD_2,...,CD_k\}$ is the set of $k$ community detection algorithms.  We perform the following steps: \\
(i) For each network, we execute $k$ algorithms present in $CD$ and obtain $k$ different community structures;\\
(ii) For each of these community structures, we compute all the scoring metrics in $SM$ separately; \\
(iii) The algorithms in $CD$ are then ranked based on the value of each of these $SM$ metrics separately, with the 
highest rank given to the highest value; \\
(iv) The community structures are further compared with the ground-truth labels of the network in
terms of each of the validation metrics in $VM$ separately; \\
(v) The algorithms are again ranked based on the values of each of the
validation metrics (highest
value/best match has the best rank); \\
(vi) Finally, we obtain Spearman's rank correlation between 
the rankings obtained for each of the scoring metrics $SM_i$ (step (iii))  and each of the ground-truth validation metric $VM_j$ (step 
(v)).

We posit that since these two types of metrics are orthogonal, and because the
validation metrics generally 
provide a stronger measure of correctness due to direct correspondence with the ground-truth structure, the
ranking of a good scoring  metric should ``match" with those of the validation metrics. We compare the
relative ranks instead of the absolute values, because the range of the values is not commensurate 
across the quantities and therefore the rank order is a more intrinsic measure.

\begin{figure}[!t]
\centering
\begin{tabular}{@{}c@{}}
\includegraphics[width=\columnwidth]{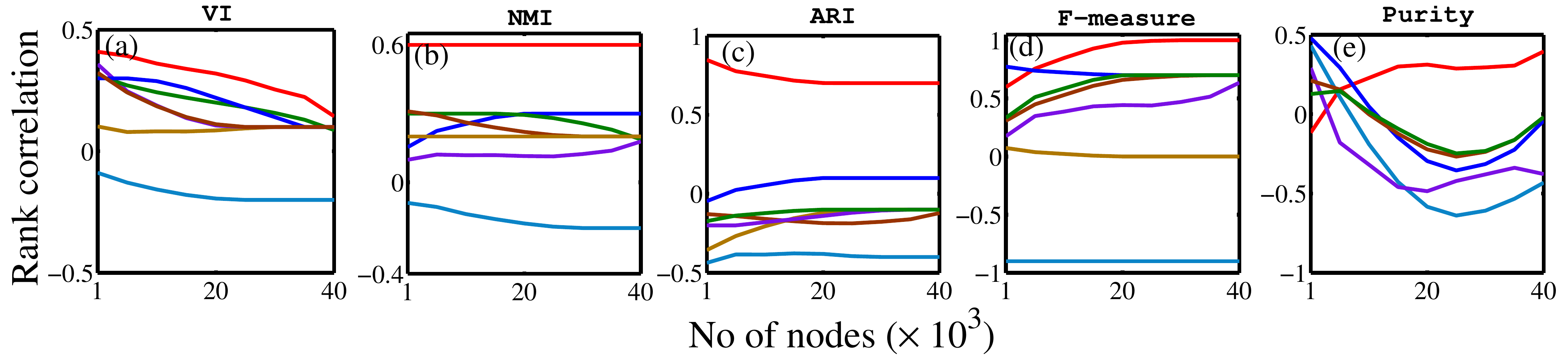}
\end{tabular}
\centering
\begin{tabular}{@{}c@{}}
\includegraphics[width=\columnwidth]{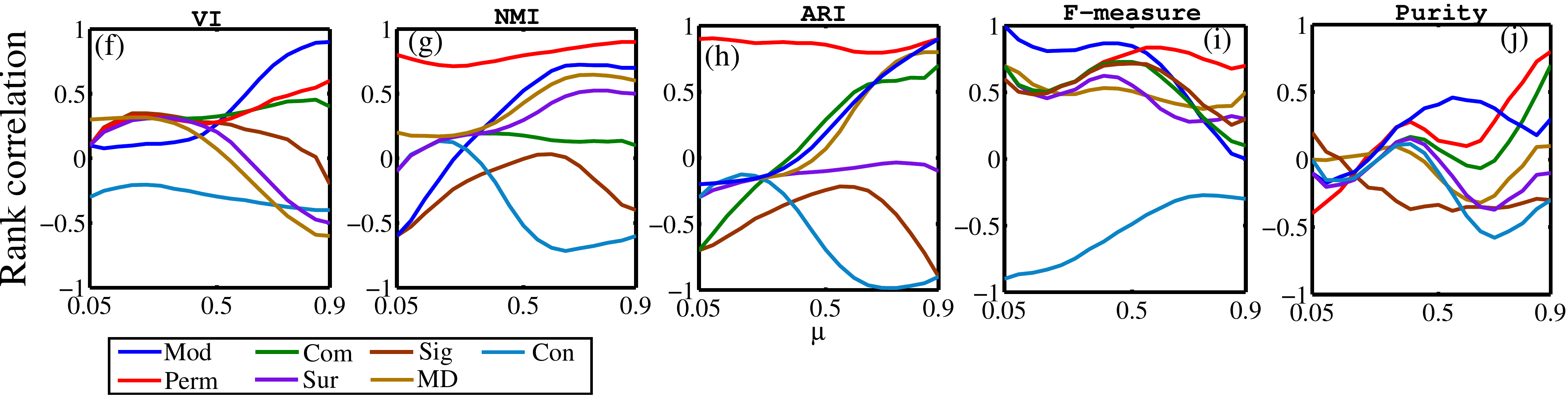}%
\end{tabular}
\caption{(Color online) Spearman's rank correlation among the results obtained from seven scoring metrics and five validation measures for LFR networks with non-overlapping community structure ((upper panel) varying the number of nodes, (lower) varying the value of $\mu$). }%
\label{lfr_non}
\end{figure}

\subsection{Comparison of non-overlapping community scoring metrics}
We compare the performance of seven state-of-the-art community scoring metrics  used as goodness measures for non-overlapping community structure: modularity (Mod), modularity density (MD),  conductance (Con), communitude (Com), asymptotic surprise (Sur), significance (Sig) and permanence (Perm). These metrics form the set $SM$ as mentioned in Section \ref{setup}. For detecting communities from the synthetic and real-world networks, we use six algorithms (representing $CD$ in Section \ref{setup}): FastGreedy \cite{newman03fast}, Louvain \cite{blondel2008}, CNM \cite{Clauset2004}, WalkTrap \cite{JGAA-124}, InfoMod, \cite{rosvall2007} and InfoMap \cite{Rosvall29012008}. To compare the output of the community detection algorithms with the ground-truth community structure, we consider five validation measures (representing $VM$ in Section \ref{setup}): variation of information (VI), normalized mutual information (NMI), adjusted rand index (ARI), F-measure (F) and purity (Pu).

Figure \ref{lfr_non} presents a comparative result of the seven scoring metrics for different LFR networks with non-overlapping community structure. In most of the cases, a general trend is observed: permanence turns out to be superior among all, which is followed by modularity; although there are few exceptions where modularity outperforms others. In most cases, communitude stands as third ranked metric, followed by modularity density and surprise. Conductance consistently performs worst among all the metrics. In few cases, we notice that while all the metrics show a decline, permanence tends to increase (Figures \ref{lfr_non}(d) and (e)) or remain consistent (Figure \ref{lfr_non}(b)). 

\begin{figure}%
\centering
\includegraphics[width=1\columnwidth]{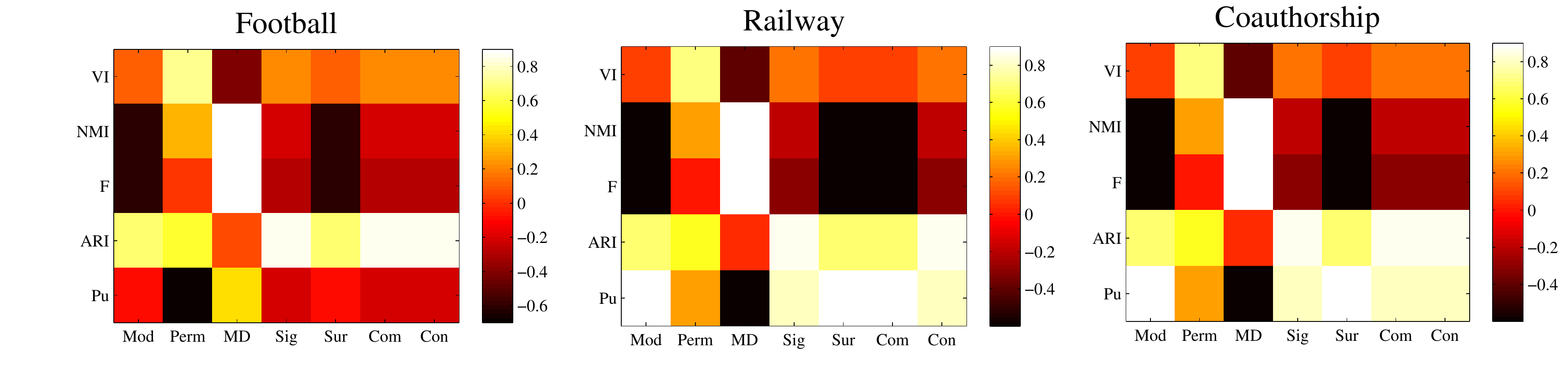}%
\caption{(Color online) Heatmap depicting the spearman's rank correlation among the results obtained from seven scoring metrics and five validation measures for real-world networks with non-overlapping community structure. }%
\label{real_non}%
\end{figure}

Figure \ref{real_non} presents a heatmap depicting the rank correlation for real-world networks. We notice that for football network, modularity density outperforms others with the average rank correlation of 0.37 (over all the validation measures, followed by permanence (0.13), significance (0.08), communitude (0.07), conductance (0.04), modularity (-0.11) and surprise (-0.11). For railway network, the result is slightly different where permanence (0.37) outperforms others. For coauthorship network which is reasonably sparse and constitutes weaker community structure, permanence (0.37) turns out to be the best, followed by significance (0.27), communitude (0.27) and conductance (0.27). In short, on average permanence performs better than others state-of-the-art metrics irrespective of the underlying network structure and validation measures.

\subsection{Comparison of overlapping community scoring metrics}
We further compare the performance of the five overlapping community scoring metrics: $Q_{ov}^{Z}$ (Equation \ref{q_z}), $Q_{ov}^{N}$ (Equation \ref{q_n}), $Q_{ov}^{S}$ (Equation \ref{eqn:shenModularity}), $Q_{ov}^{MD}$ (Equation \ref{q_md}) and flex (Equation \ref{flex}). These metrics form the set $SM$, mentioned in Section \ref{setup}. For the purpose of evaluation, we take four ground-truth based measures (representing $VM$): ONMI, Omega index, Generalized external index (GEI) and F1-score. We detect the overlapping community structure using six algorithms separately: OSLOM\footnote{{\scriptsize \url{http://www.oslom.org.}}} \cite{oslom}, EAGLE\footnote{{\scriptsize  \url{http://code.google.com/p/eaglepp/}}}
\cite{Shen20091706}, COPRA\footnote{{\scriptsize 
\url{http://www.cs.bris.ac.uk/~steve/networks/software/copra.html.}}}
\cite{Gregory1}, SLPA\footnote{{\scriptsize \url{https://sites.google.com/site/communitydetectionslpa.}}} \cite{Xie}, MOSES\footnote{{\scriptsize  \url{http://sites.google.com/site/aaronmcdaid/moses.}}}
\cite{moses} and BIGCLAM\footnote{{\scriptsize  \url{http://snap.stanford.edu}}} 
\cite{Leskovec}. These algorithms form the set $CD$. The experiment discussed in Section \ref{setup} is repeated to check which one among $SM$ highly corresponds to the results obtained from $VM$. 

Figure \ref{lfr_over} shows the results for the LFR networks by varying different parameters, i.e., $n$ and $\mu$. We also vary the parameters $O_m$ and $O_n$ (see Figure 3 in SI Text). For most of the cases, $Q_{ov}^{MD}$ seems to be the best, which is followed by flex, $Q_{ov}^{S}$, $Q_{ov}^{N}$ and $Q_{ov}^{Z}$. Most surprisingly, if we look at the trends carefully in  Figure \ref{lfr_over}, we notice that the pattern obtained by comparing with GEI is significantly different from the others. This indicates that GEI based validation measure may not be a good performance indicator for community evaluation. 

\begin{figure}[!t]
\centering
\begin{tabular}{@{}c@{}}
\includegraphics[width=\columnwidth]{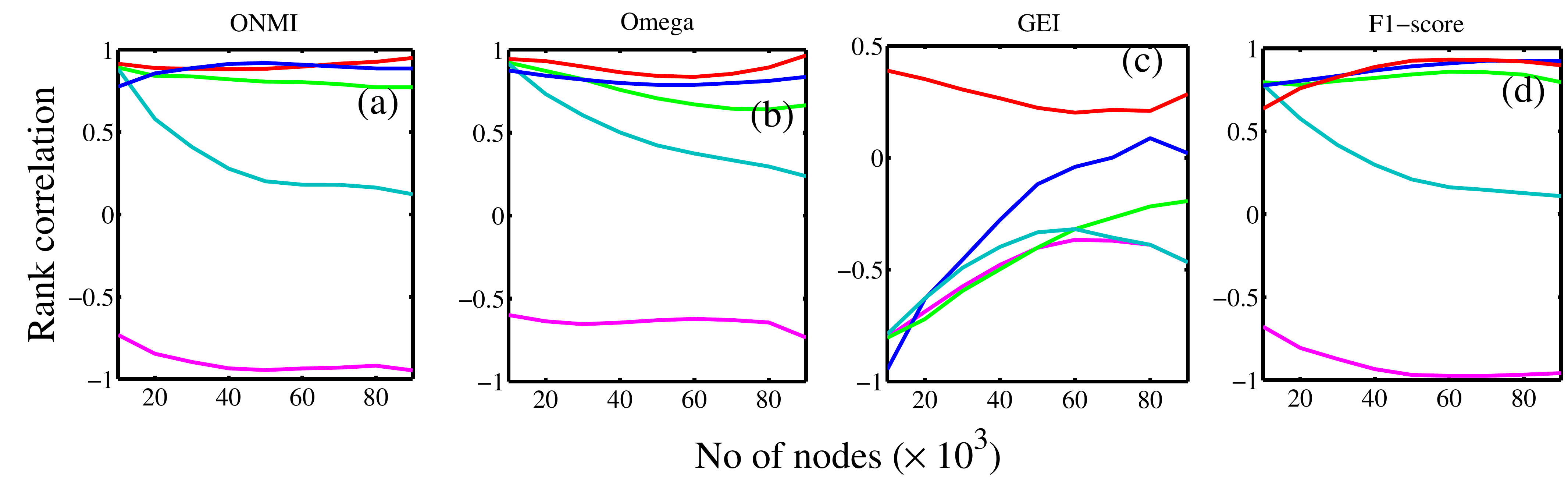}%
\end{tabular}
\centering
\begin{tabular}{@{}c@{}}
\includegraphics[width=\columnwidth]{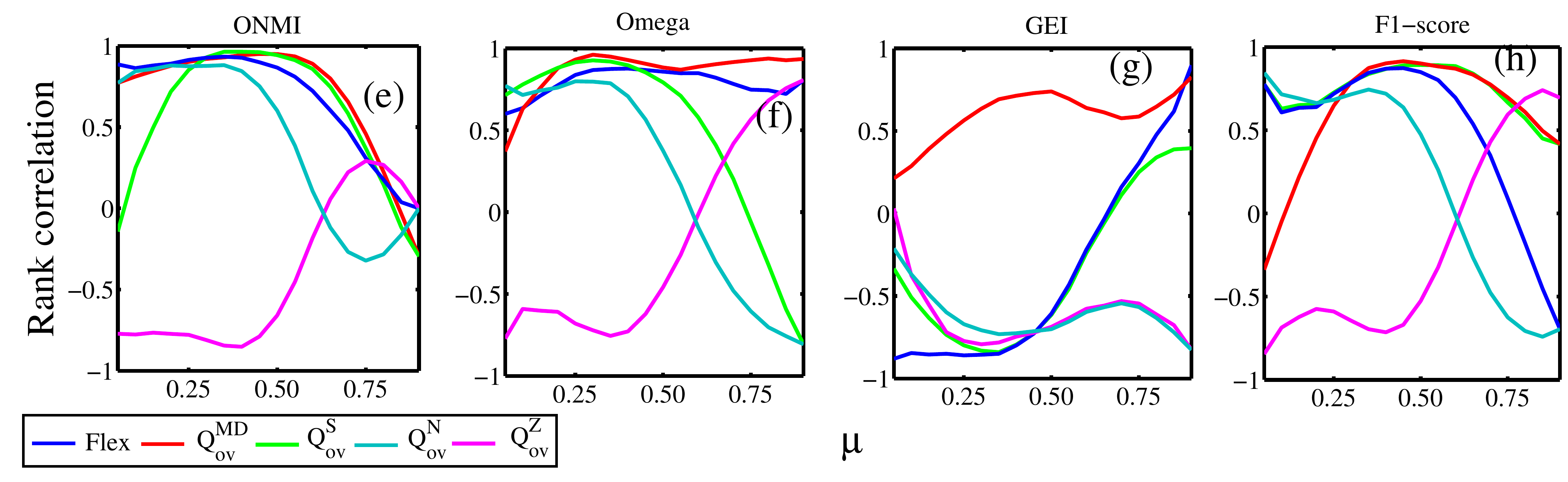}%
\end{tabular}
\caption{(Color online) Spearman's rank correlation among the results obtained from five overlapping community scoring metrics and four validation measures for LFR networks with overlapping community structure:  varying the number of nodes $n$ ($\mu=0.3$, $O_m=5$, $O_n=10\%$); mixing parameter $\mu$ ($n=10,000$, $O_m=5$, $O_n=10\%$).}
\label{lfr_over}
\end{figure}

\begin{figure}%
\centering
\includegraphics[width=1\columnwidth]{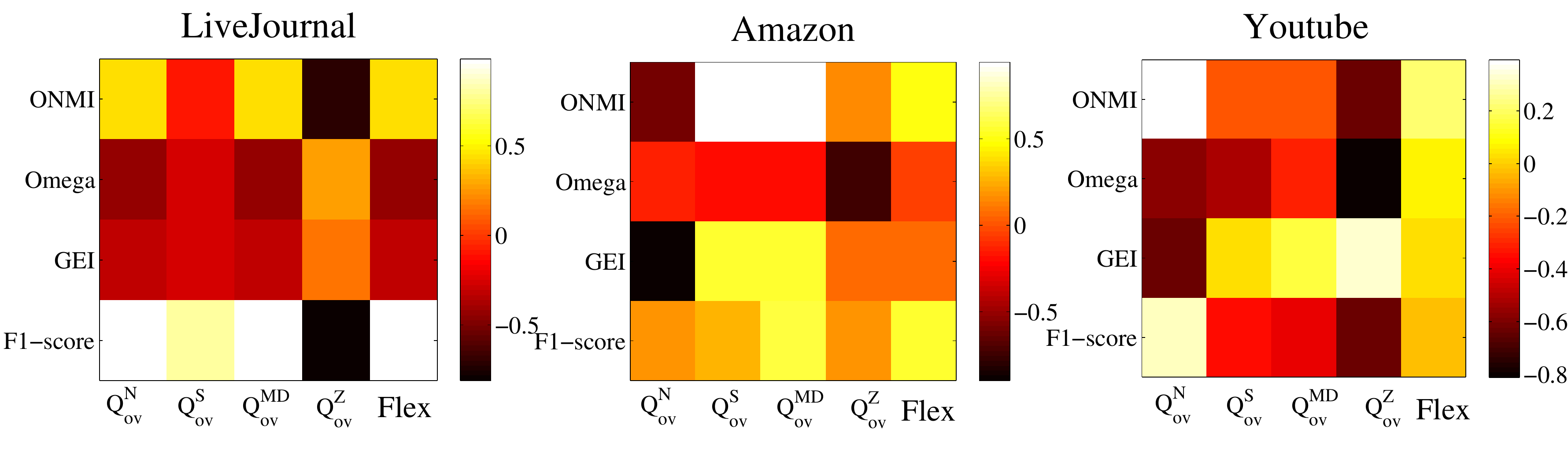}%
\caption{(Color online) Heatmap depicting the spearman's rank correlation among the results obtained from five overlapping community scoring metrics and four validation measures for real-world networks with overlapping community structure. }%
\label{real_over}%
\end{figure}

The heatmaps in Figure \ref{real_over} show the performance of the scoring metrics for real-world networks. We compute the correlation of the rank of the algorithms as discussed in Section \ref{setup}. For LiveJournal, Amazon and Youtube networks, the average correlations (over all validation measures) are reported sequentially (delimited by comma): flex (0.16, 0.26, 0.08), $Q_{ov}^Z$ (-0.27, -0.09, -0.43), $Q_{ov}^{MD}$ (0.16, 0.46, 0.19), $Q_{ov}^S$ (0.05, 0.39, -0.29) and $Q_{ov}^N$ (0.16. -0.37, -0.15). While the correlation seems to be positive (almost neutral) for flex and $Q_{ov}^{MD}$,   $Q_{ov}^Z$ and $Q_{ov}^N$ seem to be negatively correlated with the validation metrics. In short, although $Q_{ov}^{MD}$ seems to have higher correlation with the validation metrics, there is no metric which performs well on all kinds of networks. 

\section{Conclusion}\label{conc}
Despite such a vast extent of research in the detection and analysis of community structure, researchers are often in doubt while selecting an appropriate measurement metric. In this review, we attempted to understand the quality metrics pertaining to the detection of all sorts of communities. Most of these metrics are also used to evaluate the community structure. We hope that presenting all kinds of metrics together would enable the readers to understand the evolution chain of these metrics and provide them with the opportunity to select the right metric in the right context.

We observed that the most popular and widely accepted metric in the literature of community analysis is Newman-Grivan's modularity, which also lays the foundation for other metrics. Although the drawbacks of modualrity have been addressed several times, there are rare occasions where a completely new understanding of a community structure has been presented; exceptions include surprise,  significance and permanence etc. Empirical results indicated that permanence and extended modularity density ($Q_{ov}^{MD}$) are most  appropriate in measuring the quality of a community structure compared to the other competing metrics for disjoint and overlapping community detection respectively. 

An interesting question is yet to be addressed -- given a network, do we apply disjoint or overlapping community detection algorithm? There is no such metric/algorithm which is able to detect disjoint as well as  overlapping community structure depending upon the network topology without knowing the type of the underlying community strcuture. Moreover, there is very limited literature addressing the effect of various network noises on the behavior of the metric \cite{chakraborty2014permanence}. 
\cite{reichardt2006,guimera2004modularity} explored the significance of community structures only for the case of disjoint community detection which is based on the notion of modularity; although the same is yet to be explored for overlapping community structure and for other goodness metrics. Last but not the least, we believe that metrics are the core component of the community detection algorithms, and this review would have a huge implication in understanding the plethora of research in the area of community analysis.

  \newpage

{\Large{\bf Supplementary Materials}}

\section{Metrics for Non-Overlapping Communities}
\underline{\textbf{Adaptive scale modularity:}}  \cite{Twan2014Axioms} proposed six properties as axioms for community-based quality functions as follows:
\begin{itemize}
\item \textit{Permutation invariance:} This property expects that the quality of a community depends only on the weight of edges between nodes, not on the identity of nodes. Formally, a graph community quality function $Q$ is permutation invariant if for all graphs $G = (V, E)$ and all isomorphisms $f : V \rightarrow V^{\prime}$, it is the case that $Q(G, \Omega) = Q(f (G), f (\Omega));$ where $f$ is extended to graphs and communities by  $f (\Omega) = \{\{f (i) | i \in \omega\} | \omega \in \Omega\}$ and $f (V, E) = (V^{\prime} , <i, j> \rightarrow E(f^{-1}(i), f^{-1}(j)))$. \\
\item \textit{Scale invariance:} This property requires that the quality doesn't change when edge weights are scaled uniformly. Formally, a graph community quality function $Q$ is scale invariant if for all graphs $G = (V, E)$, all communities $ \omega_{i}, \omega_{j} $ of $G$ and all constants $ \alpha > 0$, $Q(G, \omega_{i} ) \leq Q(G, \omega_{j} )$ if and only if $ Q(\alpha G, \omega_{i} ) \leq Q(\alpha G, \omega_{j} )$  where $\alpha G = (V, <i, j> \rightarrow \alpha E(i, j))$ is a graph with edge weights scaled by a factor $\alpha$. \\
\item \textit{Richness:}  A graph community quality function $Q$ is rich if for all sets $V$ and all non-trivial partitions $\Omega^{*}$ of $V$, there is a graph $G = (V, E)$ such that $\Omega^{*}$ is the $Q-optimal$ communities of $V$, that is, $\Omega^{*} =argmax_{\Omega}\ Q(G, \Omega)$. \\
\item \textit{Monotonicity:} A graph community quality function $Q$ is monotonic if for all graphs $G$, all communities $\Omega$ of $G(V,E)$ and all $C-consistent$ improvements $G^{\prime}$ of $G$ it is the case that $Q(G^{\prime} , \Omega) \geq Q(G, \Omega)$. A graph $G^{\prime} = (V, E^{\prime} )$ is a C-consistent improvement of $G$ if for all nodes $i$ and $j$, $E^{\prime} (i, j) \geq E(i, j)$ whenever $i \sim_{\Omega} j$ and $E^{\prime}(i, j) \leq E(i, j)$ whenever $i \nmytilde_{\Omega} j$. \\
\item \textit{Locality:} This property demands that the contribution of a single community to the total quality should only depend on nodes in the neighborhood of that community. Formally, a quality function $Q$ is local if for all graphs
$G_{1} = (V_{1} , E_{1})$ and $G_{2} = (V_{2} , E_{2} )$ that agree on a set $V_{a}$ and its neighborhood, and for all
communities $\Omega_{a}$ , $\Omega_{a}^{\prime}$ of $V_{a}$ , $\Omega_{i}$ of $V_{1} \setminus V_{a}$ and $\Omega_{j}$ of $V_{2} \setminus V_{a}$ , if $Q(G_{1} , \Omega_{a} \cup \Omega_{i} ) \geq Q(G_{2}, \Omega_{a}^{\prime} \cup \Omega_{i} )$ then $Q(G_{1} , \Omega_{a} \cup \Omega_{j} ) \geq Q(G_{2}, \Omega_{a}^{\prime} \cup \Omega_{j} )$.\\
\item \textit{Continuity:} A quality function $Q$ is continuous if a small change in the graph leads to a small change in the quality. Formally, $Q$ is continuous if for every $\epsilon > 0 $ and every graph $G = (V, E)$ there exists a $\delta > 0$ such that for all graphs $G^{\prime} = (V, E^{\prime} )$, if $E(i, j) -\delta < E^{\prime}(i, j) < E(i, j) + \delta$ for all nodes $i$ and $j$, then $Q(G^{\prime} , \Omega) - \epsilon < Q(G, \Omega) < Q(G^{\prime} , \Omega) + \epsilon$ for all communities $\Omega$ of $G$.\\
\end{itemize}

\section{Other metrics for community detection}
\subsection{Metrics for local community detection}
Most of the  algorithms developed for detecting global community require that the graph be completely known. However, in real-world scenario, it might not be possible to collect the entire network due to several reasons, such as privacy of the data, inefficient crawling, noise during data curation. In such cases, instead of discovering global community structure of a network, people might be interested to explore the community around a particular node, i.e., the other nodes who belong to the same community/communities which the given node is a part of. This is called the local community structure. 

Although \cite{PhysRevE.72.046108} were the first who addressed the issue of incomplete network for community detection, they proposed a global community finding algorithm. Later, \cite{clauset2005finding} proposed {\bf local modularity} for local community detection.  Suppose that in the graph $G$, we have perfect knowledge of the connectivity of some set of vertices, denoted as $C$. This implies the existence of a set of vertices $U$ about which we know only their adjacencies to $C$. 
The adjacency matrix of such partially known graph is given by,
\[
   A_{ij} = 
\begin{cases}
    1,              & \text{if vertices} ~i ~\text{and} ~j ~\text{are connected, and either vertex is in} ~C\\
    0,              & \text{otherwise}
\end{cases}
\]
We restrict our consideration to those vertices in the subset of $C$ that have at least one neighbor in $U$, i.e., the vertices which make up the {\em boundary} of $C$.  Let us denote those vertices that comprise the boundary as $B$, and the boundary-adjacency matrix as
\[
   B_{ij} = 
\begin{cases}
    1,              & \text{if vertices} ~i ~\text{and} ~j ~\text{are connected, and either vertex is in} ~B\\
    0,              & \text{otherwise}
\end{cases}
\] 
Then the local modularity $R$ is defined as follows:
\begin{equation}
 R=\frac{\sum_{ij} B_{ij}\delta_{ij}}{\sum_{ij}B_{ij}}=\frac{I}{T}
\end{equation}
where $\delta_{ij}$ is $1$ when either $v_i \in B$ and $v_j \in C$ or vice versa, and is $0$ otherwise. Here, $T$ is the number of edges with one or more endpoints in $B$, while $I$ is the number of those edges with neither endpoint in $U$.

\cite{Luo:2006} further replaced  $R$ by a new local modularity, called {\bf subgraph modularity} $M$, as shown in the following:
\begin{equation}
 M=\frac{M_{in}}{M_{out}}=\frac{1/2 \sum_{ij} A_{ij} \theta_{ij}}{\sum_{ij} A_{ij} \lambda_{ij}}
\end{equation}
where $\theta_{ij}= 1$ if both node $v_i$ and node $v_j$ exist in community $C$; $0$ otherwise. And $\lambda_{ij} = 1$ means that only one, either node $v_i$ or node $v_j$, exists in community $C$.\\
\cite{Lang:2007} proposed {\bf density-isolation} as $f_{\alpha,\beta}(S)=m_S-\alpha c_S -\beta n_S$, where $\alpha$ and $\beta$ are the two parameters.   

\cite{5231879} experimented with the earlier two measures and noticed that the results usually include many outliers, i.e., the discovered communities have high recall but low accuracy, which reduces the overall community quality. They proposed to measure the {\bf community internal relation} $L_{in}$ by the average internal degree of nodes in $C$, $L_{in}=\frac{\sum_{i \in C} IK_i}{|C|}$, where $IK_i$ is the number of edges between node $i$ and nodes in $C$. Similarly, they measured the {\bf community external relation} $L_{ex}$ by the average external degree of nodes in $B$: $L_{ex}=\frac{\sum_{j \in B} EK_j}{|B|}$, where $EK_j$ is the number of connections between node $j$ and nodes outside $C$. Then they tried to maximize $L_{in}$ and minimize $L_{ex}$ at the same time, i.e., maximize $L=\frac{L_{in}}{L_{ex}}$.

\cite{Lancichinetti} proposed local-global community detection algorithm, which proposes a fitness function, as shown in the following:
\begin{equation}
 F=\frac{d_{in}^{C}}{(d_{in}^{C}+d_{ex}^{C})^{\alpha}}
\end{equation}
where $d_{in}^{C}$ and $d_{ex}^{C}$ refer to the sums of degree of the internal nodes and external nodes of community $C$, respectively, and $\alpha$ is a resolution parameter used for controlling the size of local community.

\cite{SahaHKRZ10} proposed {\bf internal density} $\frac{m_S}{n_S}$ of a set $S$ for the local community detection. \cite{Andersen:2006,Kloster:2014} used {\bf conductance} as an objective function. \cite{Tsourakakis:2013} defined {\bf edge-surplus} of a set $S$ as $f_{\alpha}(S) = m_S - \alpha \dbinom{n_S}{2}$, where $\alpha$ is the probability of existence of an edge.

\cite{WuJLZ15} criticized that most existing metrics tend to include irrelevant subgraphs in the detected local community; they referred to such irrelevant subgraphs as free riders. They systematically studied the existing goodness metrics and provided theoretical explanations on why they may cause the free rider effect. \cite{P05001} proposed new benchmark networks and evaluation measures to compare the local community detection algorithms.

\subsection{Metrics for $n$-partite networks}
A significant class of networks constitute the $n$-partite networks. The vertices of an $n$-partite network can be partitioned into $n$ disjoint sets such that no two vertices within the same set are adjacent. There are thus $n$ distinct kinds of vertices, providing a natural representation for many affiliation or interaction networks. Several variants of modularity and community quality metrics have been proposed specially for bipartite and tripartite networks. We shall review each of them in this section.

~\cite{Barber2007modularity} defined a null model appropriate for bipartite networks, and used it to define  bipartite modularity. The null model $P_{i,j}$ for a pair of communities is given by:
\begin{equation}
P_{i,j} = Ck(i)d(j)
\end{equation}
where $k(i)$ is the degree of vertices of one type and $d(j)$ is the degree of the vertices of the other type. On applying constraints we can obtain $C=1/m$. Thus, the new null model is given by,
\begin{equation}
P_{i,j} = \frac{k(i)d(j)}{m}
\end{equation}
This can be now used in the usual equation of modularity to obtain modularity for bipartite graphs as follows,
\begin{equation}
\label{eqn:barberBipartite}
Q_{b} = \frac{1}{2 m} \sum_{ij} \Bigg[ A_{ij} - P_{ij}\Bigg] \delta_{\omega_{i},\omega_{j}}
\end{equation}
~\cite{Guimer2007modularity} proposed a new metric for bipartite network. They also showed that unipartite directed networks can be viewed as bipartite networks. Thus, the same quality metric can be used. In order to define modularity for bipartite networks, certain terms are introduced as follows.
Consider the expected number of times that actor $i$  belongs to a team comprised of $m_{a}$ actors is given by $
m_{a}\frac{t_{i}}{\sum_{k}t_{k}}$, where $t_{i}$ is the total number of teams to which actor $i$ belongs. Similarly, the expected number of times that two actors $i$ and $j$ belong to team $a$ is $m_{a}(m_{a}-1)\frac{t_{i}t_{j}}{(\sum_{k}t_{k})^2}$. 
Therefore, the average number of teams in which $i$ and $j$ are expected to be together is $\frac{\sum_{a}m_{a}(m_{a}-1)}{(\sum_{a}m_{a})^2}t_{i}t_{j}$.

Using the above equation the modularity of bipartite network is given as,
\begin{equation}
\label{eqn:GuimerBipartite}
Q_{b} = \sum_{\omega \in \Omega} \Bigg[ \frac{\sum_{i \neq j \in \omega} c_{i,j}}{\sum_{a}m_{a}(m_{a}-1)} - \frac{\sum_{i \neq j \in \omega}t_{i}t_{j}}{(\sum_{a}m_{a})^2}  \Bigg]
\end{equation}
where $c_{i,j}$ is the actual number of groups in which $i$ and $j$ are together.

~\cite{Murata2009Bipartite} pointed out the weaknesses of the bipartite modularities given by ~\cite{Barber2007modularity} and ~\cite{Guimer2007modularity} in Equation \ref{eqn:barberBipartite} and Equation \ref{eqn:GuimerBipartite} respectively. Using the bipartite modularity given by Equation \ref{eqn:GuimerBipartite} two vertex types are not treated symmetrically in the definition above. The bipartite modularity focuses on the connectivities of only one vertex type (via the vertices of the other type). It is therefore not sufficient for representing the connectivities of the other vertex type. The weaknesses of bipartite modularity given by Equation \ref{eqn:barberBipartite} are two-fold: (i) the number of communities has to be searched in advance; and (ii) the numbers of communities of both vertex types have to be equal. Both weaknesses come from the bipartitioning method employed. The first weakness is fatal for practical community detection since the search for the number of communities is computationally expensive. The second weakness is also fatal for dividing real-world networks since the numbers of communities of both vertex types are often imbalanced. In order to overcome the weaknesses of the bipartite modularities, the constraint of one-to-one correspondence between communities of both types is removed in the proposed definition of bipartite modularity. One X-vertex community may corresponds to many Y-vertex communities and vice versa.

Consider a particular division of the bipartite network into X-vertex communities and Y-vertex communities, and the numbers of the communities are $L^{X}$ and $L^{Y}$,  respectively. $V^X$ and $V^Y$ are the sets of the communities of X-vertices and Y-vertices, and $V_l^X$ and $V_m^Y$ are the individual X communities that belong to the sets $(V^X = {V_1^X , ..., V_{L^X}^X})$ and $(V^Y = {V_1^Y , ..., V_{L^Y}^Y})$. Now, we can define, $e_{lm}$ (the fraction of all edges that connect vertices in $V_{l}$ to vertices in $V_{m}$ ) and $a_{l}$ as follows,
\begin{equation}
\label{eqn:murataE}
e_{lm}=\frac{1}{2m}\sum_{i \in V_{l}} \sum_{j \in V_{m}} A(i,j)
\end{equation}
\begin{equation}
\label{eqn:murataA}
a_{l}=\sum_{m}e_{lm} = \frac{1}{2m}\sum_{i \in V_{l}} \sum_{j \in V} A(i,j)
\end{equation}
The new modularity is now defined as,
\begin{equation}
\label{eqn:murata}
Q_{B} = \sum_{l} Q_{B_{l}} = \sum_{l}(e_{lm} - a_{l}a_{m}) , m = \arg\max_{k}(e_{lk})
\end{equation}
$Q_{B_{l}}$ means the deviation of the number of edges that connect the $l$-th X-vertex community and the corresponding $(m$-th$)$ Y-vertex community, from the expected number of randomly-connected edges. A larger $Q_{B_{l}}$ value means stronger correspondence between the $l$-th community and the $m$-th community.

\cite{neubauer2009towards} used the bipartite modularity given by Equation \ref{eqn:murata} by projecting the tripartite network to bipartite networks. 

\cite{murata2010detecting,murata2011tripartite} proposed modularity for tripartite networks by extending the bipartite modularity given by Equation \ref{eqn:murata}. Consider a tripartite network  G where V is composed of three types of vertices, $V^{X}, V^{Y},$ and $V^{Z}$. A hyperedge connects the triples of the vertices $(i,j,k)$, where $i \in V^{X}$ , $j \in V^{Y}$, and $k \in V^{Z}$, respectively. Suppose that $deg(i)$ is the number of hyperedges that connect to vertex $i$. Under the condition that the vertices of $V_{l}^*$, $V_{m}^*$ and $V_{n}^*$ are of different types (where * is either $X, Y,$ or $Z$) and it sums over two dimensions, such as $a_{l}^{X}, a_{m}^{Y},$ and $a_{n}^Z$. 
\begin{equation}
e_{lmm}=\frac{1}{m}\sum_{i \in V_{l}} \sum_{j \in V_{m}} \sum_{k \in V_{n}} A(i,j,k)
\end{equation}
\begin{equation}
a_{l}^{X}=\sum_{l \in V_{l}^{X}}\sum_{m}\sum_{n}e_{lmn} = \frac{1}{m}\sum_{i \in V_{l}^{X}} \sum_{j \in V^{Y}} \sum_{k \in V^{Z}} A(i,j,k)
\end{equation}
The sum over all the communities of $V^{X}$ is as follows,
\begin{equation}
\begin{split}
Q^{X} = \sum_{l} Q_{l}^{X} = \sum_{l}\sum_{m}\sum_{n}(e_{lmn} - a_{l}^{X}a_{m}^{Y}a_{n}^{Z}) \\
m ,n= \arg\max_{j,k}(e_{ljk})
\end{split}
\end{equation}
Similarly, we can define $Q^{Y}$ and $Q^{Z}$. The new tripartite modularity $Q_{t}$ is defined as the average of $Q^{X}, Q^{Y}$ and $Q^{Z}$
\begin{equation}
Q=\frac{1}{3}(Q^{X}+Q^{Y}+Q^{Z})
\end{equation}
~\cite{Suzuki2009MultiFacet} proposed a new measure which is used to extract multi-faceted community structures from bipartite networks. They extended the bipartite modularity given by Equation \ref{eqn:murata}. The modified modularity is given as follows,
\begin{equation}
Q = \frac{1}{2} \sum_{V_{k},V_{l} \in V} \frac{e_{kl}}{a_{k}} \Bigg( \frac{e_{kl}}{|E|/2} - \frac{a_{k}a_{l}}{(|E|/2)^2} \Bigg)
\end{equation}
where $e_{ij}$ and $a_{i}$ are given by Equation \ref{eqn:murataE} and Equation \ref{eqn:murataA} respectively.

\cite{Xu2015278} proposed \textbf{density-based bipartite modularity} for evaluating community structure in bipartite networks. They demonstrated that the existing modularity measurements for bipartite network community partitioning, suffer from resolution limits. To overcome this limit, they proposed a quantitative measurement for evaluating community partitioning in bipartite networks based on the concept of the average bipartite modularity degree. 
For a bipartite community $G_{i}(V_{i}^{X}, V_{i}^{Y}, E_{i})$, they defined $D(V_{i}^{X},V_{i}^{Y}) = \sum_{j \in V_{i}^{X}} \sum_{k \in V_{i}^{Y}}A(j,k)$, $D(V_{i}^{X},\bar{V_{i}^{Y}}) = \sum_{j \in V_{i}^{X}} \sum_{k \in \bar{V_{i}^{Y}}}A(j,k)$ and $D(\bar{V_{i}^{X}},V_{i}^{Y}) = \sum_{j \in \bar{V_{i}^{X}}} \sum_{k \in V_{i}^{Y}}A(j,k)$ where $\bar{V_{i}^{X}} = V^{X} - V_{i}^{X}$ and $V_{i}^{X} \cap \bar{V_{i}^{X}} = \phi$. The bipartite density $Q_{D}(G_{i})$ of bipartite community $G_{i}$ is defined as,
\begin{equation}
Q_{D}(G_{i}) = D_{in}(G_{i}) - D_{out}(G_{i})
\end{equation}
where $D_{in}(G_{i})$ and $D_{out}(G_{i})$ are the average inner and outer degree of the community $G_{i}$ respectively. Therefore, the  bipartite density of community $G_{i}(V_{i}^{X},V_{i}^{Y},E_{i})$ can be formulated as follows:
\begin{equation}
Q_{D}(G_{i}) = \frac{D(V_{i}^{X},V_{i}^{Y}) - D(V_{i}^{X},\bar{V_{i}^{Y}}) - D(\bar{V_{i}^{X}},V_{i}^{Y})}{|V_{i}^{X}| \times |V_{i}^{X}|}
\end{equation}
Here, the intuitive idea is that $Q_{D}(G_{i})$ should be as large as possible for a valid community in bipartite networks. Next, the bipartite density-based modularity of a partitioning scheme $P$ is defined as the summation of bipartite densities overall communities $G_{i}$ for $i=1,2, \cdots, |P|$. Let $Q_{D}(P)$ denote the bipartite density-based modularity of a partition scheme $P$, which divides the bipartite network $G$ into communities $G_{i}, \cdots, G_{|P|}$ ; therefore,  $Q_{D}(P)$  can be calculated as follows:
\begin{equation}
  Q_{D}(P) = \sum_{i=1}^{c} Q(G_{i}) = \sum_{i=1}^{c}  \frac{D(V_{i}^{X},V_{i}^{Y}) - D(V_{i}^{X},\bar{V_{i}^{Y}}) - D(\bar{V_{i}^{X}},V_{i}^{Y})}{|V_{i}^{X}| \times |V_{i}^{X}|}
\end{equation}
The larger the value $Q_{D}(P)$ is, the more accurate the partitioning scheme $P$ becomes. 

\cite{Max-Intensity} proposed a vertex metric \textbf{intensity score} to measure the quality of communities in a advertiser-keyword network modeled as a bipartite graph. Given a weighted bipartite advertiser-keyword graph $G=(A,K,E)$, where $A$ is the set of advertisers, $K$ is the set of keywords and $E$ is the weighted edge such that $A \cap K = \phi$ and $E \subseteq V \times K$. Let $w_{ij}$ denote the weight of an edge between vertex $i \in A$ and vertex $j \in K$. Let $C(A_{C},K_{C},E_{C})$ be one community in the advertiser-keyword graph $G$. They defined the \textit{homogeneous neighborhood} of a vertex $u$ as a set $N(u) = \{v|(u,t) \in E_{C} \wedge (t,v) \in E_{C}, t \neq \phi, u \neq v \}$. The competition coefficient of a vertex $i \in A_{C}$ is defined as follows:
\[
   cc_{i} = 
\begin{cases}
    \frac{\sum_{j}\sum_{k} w_{jk}}{|N(i)|}, & \text{if} \quad N(i) \neq \phi \\
    0,              & \text{if} \quad N(i) = \phi
\end{cases}
\]
The intensity score $I_{i}$ of a vertex $i$ in community  $C(A_{C},K_{C},E_{C})$ is given by
\begin{equation}
	I_{i} = cc_{i} + \lambda \Bigg( \sum_{i \in A_{C} \wedge j \in K_{C}} w_{ij} - \max_{c^{\prime}} \sum_{i \in A_{C} \wedge j^{\prime} \in K_{C^{\prime}}} w_{ij^{\prime}} \Bigg)
\end{equation}
where $ \lambda \geq 0 $ is a tuning parameter. The objective is to partition the vertices of $G$ into subsets so as to maximize the intensity score in each detected community. The goal function is,
\begin{equation}
f_{obj} = \max \sum_{A_{C} \subseteq G} \sum_{i \in A_{C}} I_{i}
\end{equation}
 \subsection{Community quality metrics for anti-community detection}
So far we discussed about metrics to find assortative communities. However, there has been an interest to find anti-communities. In an anti-community, vertices have most of their connections outside their group and have no or fewer connections with the members within the same group. ~\cite{estrada2005Bipartivity,Newman2006eigenvectors,Newman05062007} did some of the initial works to propose ``disassortative'' communities or communities with networks possessing bipartite structure. However most of these methods either modify the existing definition of modularity or propose a new algorithm for anti-community detection. 

Recently, ~\cite{Chen2014293} proposed a measure, \textbf{anti-modularity} to find anti-communities. Consider $B=[b_{ij}] = A^TA$. If $v_{1}$ and $v_{2}$ are the vertices in the same anti-community, then $b_{ij} = \sum_{k=1}^{N}a_{ik}a_{kj}$ is the number of paths between $v_{i}$ and $v_{j}$ passing through a third vertex. The expected number of paths between $v_{i}$ and $v_{j}$ passing through a third vertex is next estimated. For a vertex $v_{k} \in V$ the probability that it appears in the immediate neighbor of  $v_{i}$ and $v_{j}$  is $\frac{d(i)d(j)}{N^2}$. Thus, for all $N$ vertices the expectation of the number of vertices is  $N.\frac{d(i)d(j)}{N^2} =  \frac{d(i)d(j)}{N}$. Thus, for an anti-modularity partition $\Omega^{prime} = {\omega_{1}^{\prime}, \omega_{2}^{\prime}, \cdots, \omega_{|\Omega|}^{\prime}}$, the anti-modularity is defined as:
\begin{equation}
Q_{anti} = \frac{1}{N} \sum_{\omega^{\prime} \in \Omega^{\prime}} \sum_{v_{i}, v_{j} \in V_{\omega^{\prime}}} \Bigg( \sum_{k=1}^{N} a_{ik}a_{jk} - \frac{d(i)d(j)}{N} \Bigg)
\end{equation}
They demonstrated via experiments that optimizing anti-modularity is reasonable and reliable measure of anti-community partitioning.

\begin{table}[!th]
\centering

\tbl{Metrics for non-overlapping community detection.\label{tab:otherMetrics}}{
\scalebox{0.7}{
\begin{tabular}{|l|l|}
\hline
Metric            & Expression \\
\hline
Internal density & $\frac{|E_{\omega}^{in}|}{|\omega|(\omega-1)/2}$ \\
\hline
Edge inside & $|E_{\omega}^{in}|$ \\
\hline
Average degree & $2|E_{\omega}^{in}|/|\omega|$ \\
\hline
Fraction over median degree (FOMD) & $\frac{|{u:u\in \omega,|{(u,v):v\in \omega}|> d_m}|}{|\omega|}$  \\                    
\hline
Triangle Participation Ratio (TPR) & $\frac{|{u:u\in \omega,\{v,w\in \omega,(u,v)\in E,(u,w)\in E,(v,w)\in E\} \neq \phi}|}{|\omega|}$ \\
\hline
Expansion & $|E_{\omega}^{out}|/|\omega|$ \\
\hline
Cut Ratio & $|E_{\omega}^{out}|/|\omega|(N−|\omega|)$ \\
\hline
Conductance & $\frac{|E_{\omega}^{out}|}{2|E_{\omega}^{in}|+|E_{\omega}^{out}|}$ \\
\hline
Normalized Cut & $\frac{|E_{\omega}^{out}|}{2|E_{\omega}^{in}|+|E_{\omega}^{out}|} + \frac{|E_{\omega}^{out}|}{2(m - |E_{\omega}^{in}|)+|E_{\omega}^{out}|}$ \\
\hline
Maximum-ODF (Out Degree Fraction) & $ max_{u\in \omega} \frac{|{(u,v)\in E:v \notin \omega}|}{d(u)}$ \\
\hline
Average-ODF & $\frac{1}{|\omega|} \sum_{u \in \omega} \frac{|{(u,v)\in E:v \notin \omega}|}{d(u)}$ \\
\hline
Flake-ODF & $\frac{|{u:u\in \omega,|{(u,v)\in E:v\in \omega}|<d(u)/2}|}{|\omega|}$ \\
\hline
Modularity & $\sum_{\omega \in \Omega} \Bigg[ \frac{|E_{\omega}^{in}|}{m} - \Bigg(\frac{ |E_{\omega}^{in}+ E_{\omega}^{out}|}{2m}\Bigg)^{2} \Bigg]$ \\
\hline
Separability & $|E_{\omega}^{in}|/|E_{\omega}^{out}|$ \\
\hline
Density & $2|E_{\omega}^{in}|/\omega(\omega-1)$ \\
\hline
Cohesiveness & $min_{\omega^{'} \subset \omega} \phi(\omega^{'})$ , where $\phi(\omega^{'})$ is the conductance of $\omega^{'}$ measured in the induced subgraph by $\omega$. \\
\hline
Volume & $\sum_{u\in \omega} d(u)$ \\
\hline
Edges cut &  $|E_{\omega}^{out}|$ \\
\hline
Weighted Modularity & $\frac{1}{2 |W|} \sum_{ij} \Bigg[ W_{ij} - \frac{s(i)s(j)}{2|W|} \Bigg] \delta_{\omega_{i},\omega_{j}}$ \\
\hline
Modularity for Directed Graphs & $\frac{1}{ m} \sum_{ij} \Bigg[ A_{ij} - \frac{d(i)^{out} d(j)^{in}}{m} \Bigg] \delta_{\omega_{i},\omega_{j}}$ \\
\hline
Generalized Modularity &  $\frac{1}{ |W|} \sum_{ij} \Bigg[ W_{ij} - \frac{s_{i}^{out}s_{j}^{in}}{|W|} \Bigg] \delta_{\omega_{i},\omega_{j}}$ \\
\hline
Coverage-based modularity & $\frac{\sum_{\omega \in \Omega} \frac{|E_{\omega}^{in}|}{m}}{\sum_{\omega \in \Omega} \frac{|E_{\omega}^{in}+E_{\omega}^{out}|}{2m}}$ \\
\hline
Similarity-based modularity & $\sum_{\omega \in \Omega} \Bigg[ \frac{\sum_{i} \sum_{j} S(i,j) \delta(i,\omega) \delta_(j,\omega) }{ \sum_{i} \sum_{j} S(i,j)} - \Bigg( \frac{\sum_{i} \sum_{j} S(i,j) \delta(i,\omega)}{\sum_{i} \sum_{j} S(i,j)} \Bigg)^{2} \Bigg]$ \\
\hline
Motif modularity & $\Bigg[ \frac{\sum_{ijk} A_{ij}(\omega) A_{jk}(\omega) A_{ki}(\omega)} {\sum_{ijk} A_{ij} A_{jk} A_{ki}} - \frac{\sum_{ijk} n_{ij}(\omega) n_{jk}(\omega) n_{ki}(\omega)} {\sum_{ijk} n_{ij} n_{jk} n_{ki}} \Bigg]$ \\
\hline
Max-Min modularity & $\sum_{ij} \Bigg[ \frac{1}{2m} \Bigg( A_{ij} -\frac{d(i)d(j)}{2m} \Bigg) - \frac{1}{2m^{\prime}} \Bigg( A_{ij}^{\prime} -\frac{d(i)^{\prime}d(j)^{\prime}}{2m^{\prime}}  \Bigg) \Bigg]\delta_{\omega_{i},\omega_{j}}$ \\
\hline
Influence-based modularity & $\sum_{ij} \Bigg[ P_{ij} - \bar{P_{ij}}\Bigg] \delta_{\omega_{i}, \omega_{j}}$ \\
\hline
Diffusion-based modularity & $\sum_{ij} \Bigg[ L_{ij} - E(L_{ij}) \Bigg] \delta_{\omega_{i},\omega_{j}}$ \\
\hline
Dist-modularity & $\frac{1}{2 m} \sum_{ij} \Bigg[ A_{ij} - P_{ij}^{Dist} \Bigg] \delta_{\omega_{i},\omega_{j}}$ \\
\hline
Local modularity & $\sum_{\omega \in \Omega} \bigg[ \frac{|E^{in}_{\omega}|}{L_{{\omega}n}} - \frac{|E^{in}_{\omega}| |E^{out}_{\omega}|}{L_{{\omega}n}^2} \bigg]$ \\
\hline
Modularity Density & $\sum_{\omega \in \Omega} d(G_{\omega})$\\
\hline
Modularity with Resolution Parameter & $\sum_{\omega \in \Omega} \Bigg[ \frac{|E_{\omega}^{in}|+|V_{\omega}|r}{|E|+|V|r} - \Bigg(\frac{ |E_{\omega}^{in}+ E_{\omega}^{out}|+|V_{\omega}|r}{2|E|+|V|r}\Bigg)^{2} \Bigg]$ \\
\hline
Modularity with Diameter & $\sum_{\omega \in \Omega} \Bigg[ \frac{|E_{\omega}^{in}|}{D_{\omega}} - \Bigg(\frac{ |E_{\omega}^{in}+ E_{\omega}^{out}|}{2|E|}\Bigg)^{2}*\frac{1}{\tilde{D_{\omega}}} \Bigg]$ \\
\hline
Modularity Density & $Q_{ud}-SP$ where $SP = \sum_{\omega_{i} \in \Omega} \Bigg[ \sum_{\substack{\omega_{j} \in \Omega \\ \omega_{j} \neq \omega_{i}}} \frac{|E_{\omega_{i},\omega_{j}}|}{2m}\Bigg]$ \\
\hline
Z-Modularity & $\sum_{\omega \in \Omega} {\Bigg( \frac{D_{\omega}}{2m} \Bigg)}^{2}$ \\
\hline
Community Score & $\sum_{\omega \in \Omega} score(\omega)$, $score(\omega) = M(\omega)v_{\omega}$, $M(\omega) = \frac{\sum_{i \in \omega}(\mu_{i})^{r}}{N_{\omega}}$ \\
\hline
SPart & $\frac{1}{|\Omega|} \sum_{i \in \Omega} SComm(\omega_{i}) \frac{ \nu(\omega_{i})}{|\omega_{i}|}$ \\
	 &  $SNode(v) = \frac{d(v)^{in} - d(v)^{out}}{|\omega|}$ \\
	 &  $SComm(\omega) = \sum_{v \in \omega} \Bigg [ SNode(v) + \frac{1}{2} \sum_{\substack{w \in \omega \\ A_{vw}=1} } SNode(w) \Bigg]$\\
\hline
Significance & $- \log Pr(\Omega) = \sum_{c} {{n_\omega} \choose 2} D(p_\omega || p)$\\
			& $Pr(\Omega) = \prod_{\omega} exp (- {{n_{\omega}}\choose{2}} D (p_{c}||p))$ \\ 
\hline
Permanence & $ \Bigg[\frac{I(v)}{E_{max}(v)}\times \frac{1}{d(v)}\Bigg] - \Bigg[ 1 - c_{in}(v) \Bigg]$ \\
\hline
Surprise & $ \sum_{j=p}^{Min(M,|E|)} \frac{ {{M} \choose{j}}{{F-M}\choose{|E|-j}}}{{{F}\choose{|E|}}}$ \\
\hline
Expected nodes & $\mu_{G}(|\omega|) = \sum_{u \in V} 1 - \frac{{{2|E|-b(u)} \choose {2|\omega|}}}{{{2|E|} \choose {2|\omega|}}}$\\
				& $Q_{in}(\omega) = \frac{\mu_{G}(|\omega|)-V_{in}(\omega)}{\mu_{G}(|\omega|)}$ \\
				& $Q_{ext}(\omega) = \min \Bigg( 0, \frac{|V_{out}(\omega)| - \mu_{G \setminus \omega} (\bar{d}(\omega)/2)  }{\mu_{G \setminus \omega} (\bar{d}(\omega)/2)} \Bigg)$ \\
				& $Q_{\omega}= 2 \frac{|\omega|Q_{in}(\omega)+|\omega_{out}|Q_{ext}(\omega)}{|\omega|+|\omega_{out}|}$ \\
				& $\frac{\sum_{\omega \in \Omega} |\omega| Q(\omega) }{|E|}$ \\
\hline
Communitude & $\frac{\frac{|E_{\omega}^{in}|}{m} - (\frac{|E_{\omega}^{in}+E_{\omega}^{out}|}{2m})^2}{(\frac{|E_{\omega}^{in}+E_{\omega}^{out}|}{2m})^2(1-(\frac{|E_{\omega}^{in}+E_{\omega}^{out}|}{2m})^2)}$ \\
\hline
\end{tabular}}}
\end{table}

\begin{table}[!th]
\centering
\tbl{Metrics for Overlapping Community Detection.\label{tab:otherMetrics}}
{
\scalebox{0.8}{
\begin{tabular}{|l|l|}
\hline
Metric  & Expression \\
\hline
Modularity & $\sum_{\psi \in \Psi} \Bigg[ \frac{|E_{\psi}^{in}|}{m} - \Bigg(\frac{2|E_{\psi}^{in}|+|E_{\psi}^{out}| }{2m}\Bigg)^{2} \Bigg]$ \\
\hline
Fuzzy Modularity & $\frac{1}{2m} \sum_{ij} \Bigg[ A_{ij} - \frac{d(i)d(j)}{2m} \Bigg] s_{ij}$ \\
\hline
Modularity~\cite{shen2009overlap} & $\frac{1}{2m}  \sum_{ij} \Bigg[ A_{ij} - \frac{d(i)d(j)}{2m} \Bigg] \alpha_{i\psi} \alpha_{j\psi}$ \\
\hline
Modularity~\cite{Shen20091706} & $\frac{1}{2 m} \sum_{\psi \in \Psi} \sum_{ij} \Bigg[ A_{ij} - \frac{d(i)d(j)}{2m} \Bigg] \frac{1}{O_{i}O_{j}}$ \\
\hline         
Modularity~\cite{nicosia2009Extension} & $\frac{1}{2m} \sum_{\psi \in \Psi} \sum_{ij} \Bigg[ A_{ij} F \Big( \alpha_{i\psi},\alpha_{j\psi} \Big) - \frac{d(i)d(j) \Bigg( \sum_{v \in V} F \Big(\alpha_{v\psi}, \alpha_{j\psi} \Big) \Bigg) \Bigg( \sum_{v \in V} F \Big(\alpha_{i\psi}, \alpha_{v\psi}\Big) \Bigg)}  {2mN^{2}} \Bigg] $ \\ 
\hline 
Fitness Function & $\frac{|E_{in}|^{\psi}}{(|E_{in}|^{\psi}+|E_{out}|^{\psi})^\alpha}$ \\
\hline
Modularity~\cite{lazar2010modularity} & $\frac{|E_{\psi}^{in}|+|E_{\psi}^{out}|}{N_{\psi}(N_{\psi}-1)/2} \frac{1}{N_{\psi}} \sum_{i \in N_{\psi}} \frac{ \sum_{j \in N_{\psi} i \neq j} A_{ij} - \sum_{j \notin N_{\psi}} A_{ij} } {d(i)s_{i}}$ \\
\hline
Modularity~\cite{chen2010Overlapping} & $\frac{1}{2m} \sum_{\psi \in \Psi} \sum_{ij} \Bigg[ A_{ij} - \frac{d(i)d(j)}{2m} \Bigg] \alpha_{i\psi} \alpha_{j\psi}$ \\
\hline
Modified Fitness Function & $ |E_{in}|^{\psi}+|E_{out}|^{\psi})^\alpha $ \\
\hline
Overlapping Modularity Density & $ \sum_{\psi \in \Psi} \Bigg[ \frac{|E_{\psi}^{in}|}{m}d_{\psi} - {\Bigg( \frac{ 2|E_{\psi}^{in}|+|E_{\psi}^{out}| }{2m} d_{\psi} \Bigg)}^{2} - \sum_{\psi^{'} \in \Psi, \psi \neq \psi^{'}} \frac{|E_{\psi,\psi^{'}}|} {2m} d_{\psi,\psi^{'}} \Bigg],$ \\
\hline
Flex & $LC(i,\psi) = \lambda* \triangle(i,\psi) + (1-\lambda)*N(i,\psi) - \kappa * \wedge(i,\psi)$ \\
	&	$CC(\psi) = \sum_{i \in \psi} LC(i,\psi) - \frac{{N_{\psi}}^{\gamma}}{N}$ \\
	&	$\frac{1}{N} \sum_{\psi \in \Psi} CC(\psi)$ \\
\hline
\end{tabular}}}
\end{table}

\begin{table}[!th]
\centering
\tbl{Metrics for Non-Overlapping Community Evaluation.\label{tab:otherMetrics}}
{

\begin{tabular}{|l|l|}
\hline
Metric  & Expression \\
\hline
Purity & $\frac{1}{N}\sum_k \underset{j}{\max} |\omega_k,c_j|$ \\
\hline
F-Measure & $\frac{2\cdot Purity(\Omega,C) \cdot Purity(C,\Omega)}{Purity(\Omega,C) + Purity(C,\Omega)}$ \\
\hline
Rand Index & $\frac{TP+TN}{TP+FP+FN+TN}$ \\
\hline
$F_{\beta}$ & $\frac{(\beta^2+1)PR}{\beta^2P+R}; P=\frac{TP}{TP+FP}; R=\frac{TP}{TP+FN};$ \\
\hline
Adjusted Rand Index & $ \frac{\sum_{ij}\dbinom{N_{\omega_ic_j}}{2} - \sum_i\dbinom{N_{\omega_i}}{2} \sum_j \dbinom{N_{c_j}}{2} / \dbinom{N}{2}}{\frac{1}{2} \bigg(\sum_i\dbinom{N_{\omega_i}}{2} + \sum_j \dbinom{N_{c_j}}{2} \bigg) - \sum_i\dbinom{N_{\omega_i}}{2} \sum_j \dbinom{N_{c_j}}{2} / \dbinom{N}{2}}$ \\
\hline
Normalized Mutual Information & $\frac{I(\sigma,C)}	{[H(\sigma)+H(C)]/2}$ \\
                              & $I(\Omega,C)=\sum_k\sum_j \frac{|\omega_k \cap c_j|}{N}\ log\frac{{N|\omega_k \cap c_j|}}{|\omega_k||c_j|}$ \\
\hline
Entropy & $H(\Omega)=-\sum_k \frac{|\omega_k|}{N}\ log\frac{\omega_k}{N}$ \\
\hline
Variation of Information &  $\sum_{i,j}r_{ij}[ \log (r_{ij}/p_i) + \log (r_{ij}/q_j)]$ \\
             & $H(\Omega)+H(C)-2I(\Omega,C)$ \\
\hline
Modified Purity &  $\sum_i \sum_{u\in \omega_i} \frac{w_u}{w} Purity (u,\Omega,C)$ \\
\hline
Modified ARI & $ \frac{\sum_{ij}W(\omega_i\cap W(c_j) - \sum_j W(\omega_i)W(c_j)/W(V)} {\frac{1}{2} \big( \sum_i W(\omega_i) + \sum_j W(c_j) \big) - \sum_i W(\omega_i) \sum_j W(c_j) / W(s) }$ \\
\hline
\end{tabular}}
\end{table}

\begin{table}[!th]
\centering
\tbl{Metrics for overlapping community evaluation.\label{tab:otherMetrics}}
{
\scalebox{0.85}{
\begin{tabular}{|l|l|}
\hline
Metric            & Expression \\
\hline
Overlapping Normalized Mutual Information & $ONMI(X|Y) = 1 - [H(X|Y)+ H(Y|X)]/2$ \\  
\hline
Omega Index & $\frac{Omega_u(\Psi,C)-Omega_e(\Psi,C)}{1-Omega_e(\Psi,C)}$ \\
			& $\frac{1}{M}\sum_{j=1}{max(|\Psi|,|C|)} |t_j(\psi_i) \cap t_j(c_j)|$ \\
			& $\frac{1}{M^2}\sum_{j=1}{max(|\Psi|,|C|)} |t_j(\psi_i)|\cdot|t_j(c_j)|$ \\
\hline
Generalized External Index &  $a_G(i,j)=min\{\alpha_{\Psi}(i,j),\alpha_{C}(i,j)\} +  min\{\beta_{\Psi}(i),\beta_{C}(i)\} + min\{\beta_{\Psi}(j),\beta_{C}(j)\}$ \\
 & $d_G(i,j)=abs[\alpha_{\Psi}(i,j)-\alpha_{C}(i,j)] +  abs[\beta_{\Psi}(i)-\beta_{C}(i)] + abs[\beta_{\Psi}(j)-\beta_{C}(j)]$ \\
  & $GEI(\Psi,C)=\frac{a_G}{a_G+d_G}$ \\
\hline
F1-Score & $\frac{1}{2} (\frac{1}{|\Psi|}  \sum_{\psi_i \in \Psi} F1(\psi_i,C_{g(i)}) + \frac{1}{|C|}  \sum_{c_i \in C} F1(\Psi_{g^{'}(i)},c_i ))$ \\
\hline
\end{tabular}}}
\end{table}

\subsection{Community quality metrics for multiplex networks}
A multiplex is a finite set of $m$ graphs, $G = \{G_{1}, \cdots, G_{m}\}$, where every graph $G_{i} = (V, E_{i})$ has a distinct edge set $E_{i} \subseteq V \times V$. Multiplex networks have recently caught attention in the field of community detection. The existing quality metrics discussed so far for monoplex networks may not give the desired communities in case of multiplex networks. In this section, we discuss the different community detection quality measures designed to measure the quality of a community in a multiplex network.

~\cite{tang2009multilayer} built along the lines of the community measure developed by Newman and Girvan. The modularity function by Newman and Girvan measures how different a monoplex-communities are from a random graph. Given a fixed partition on the vertex set, the modularity on each of the $m$ graphs in the multiplex differs. Therefore a good multiplex-communities suggests that all the monoplex-communities in the graphs have high modularity.

To quantify this concept, ~\cite{tang2009multilayer} claimed that if there exists latent communities in the multiplex, a subset of the graphs in the multiplex, $\mathcal{G^{\prime}} \subset \mathcal{G}$ has sufficient information to find these communities. If the hypothesis is true, then the communities detected from $\mathcal{G^{\prime}}$ should reflect high modularity on the rest of the graphs in the multiplex. 

In the language of machine learning, let us pick a random graph $G \in \mathcal{G}$ as the test data and let $ \mathcal{G^{\prime}} = \mathcal{G} \in G$ be the training data. The multiplex-partition $P$ yielded from a community detection algorithm on $\mathcal{G^{\prime}}$ is evaluated with the modularity function on the test data $G$. $P$ is a good multiplex-partition if the modularity of partition $P$ on the graph $G$ is maximized. This extends the modularity metric for multiplex.

\cite{berlingerio2011multiDimensional} proposed a local metric, \textbf{redundancy} as a measure to compute the quality of a multiplex community.  Consider each graph in a multiplex as an independent mode of communication between the members, e.g., email, telephone, postal, etc. A high quality community should resume high information flow amongst its members when one of the communication modes fails. 

Let $W \subseteq V$ be the set of vertices in a multiplex-community and $P \subseteq W \times W$ be the set of vertex pairs in $W$ that are adjacent in $\geq 1$ relationship. The set of redundant vertex pairs are $P^{\prime} \subseteq P$ where vertex pairs in $W$ that are adjacent in $\geq 2$ relationships. The redundancy of $W$ is determined by:
\begin{equation}
\frac{1}{\mathcal{|G|} \times \mathcal{|P|}} \sum_{G^{\prime} \in \mathcal{G}} \sum_{\{u,v\} \in P^{\prime}} \delta(u,v,E_{i})
\end{equation}
where $\delta(u,v,E_{i}) = 1 \quad if \quad \{u,v\} \in E_{i} \quad (0, \quad otherwise)$. The metric essentially counts the number of edges in the multiplex-community where their corresponding vertex pairs are adjacent in two or more graphs. The sum is normalized by the theoretical maximum number of edges between all adjacent vertex pairs, i.e., $(\mathcal{|G|} \times \mathcal{|P|})$. The quality of a multiplex-community is determined by how identical the subgraphs (induced by the vertices of the multiplex-community) are across the graphs in the multiplex. Thus the redundancy does not depend on the number of edges in the multiplex-community, i.e., not a necessary condition to its quality. This can lead to an unusual idea that a community can be low in density. For instance a cycle of overlapping edges form a ``community'' of equal quality as a complete clique of overlapping edges. 

\cite{aberer2012multiLayer} proposed a vertex based community quality measure, \textbf{Cross-Layer Edge Clustering Coefficient (CLECC)}. A high edge clustering coefficient implies that there are many common neighbors between the vertex pair, thus suggesting that the two vertices should belong to the same community. Given a parameter $\alpha$, the MIN-Neighbors of vertex $v$, $N (v, \alpha)$ are the set of vertices that are adjacent to $v$ in at least $\alpha$ graphs. The Cross-Layer Edge Clustering Coefficient of two vertices $u, v \in V$ measures the ratio of their common neighbors to all their neighbors. 
\begin{equation}
CLECC(u,v,\alpha) = \frac{|N(u, \alpha) \cap  N(v, \alpha)|}{|N(u, \alpha) \cup N(v, \alpha) \setminus \{ u,v\}|}
\end{equation}
A pair of vertices in a multiplex of social networks with low CLECC  suggests that the individuals do not share a common clique of friends through at least $\alpha$ social networks. Therefore it is unlikely that they form a community together.

\cite{Loe201529} studied the different methods proposed to detect communities in \textit{multiplex networks} and performed a comparative analysis of the same.
\subsection{Community detection in signed networks}
Community Detection has been significantly under-explored for networks with positive and negative links as compared to unsigned ones. Trying to fill this gap, \cite{esmailian2015community} measured the quality of partitions by introducing a Map Equation for signed networks. It is based  on the assumption that negative relations weaken positive flow from a node towards a community, and thus, external (internal) negative ties increase the probability of staying inside (escaping from) a community.
\section{Metrics for Community Evaluation}
\subsection{Issues with traditional measures}
\begin{figure}
 \centering
 \includegraphics[width=\columnwidth]{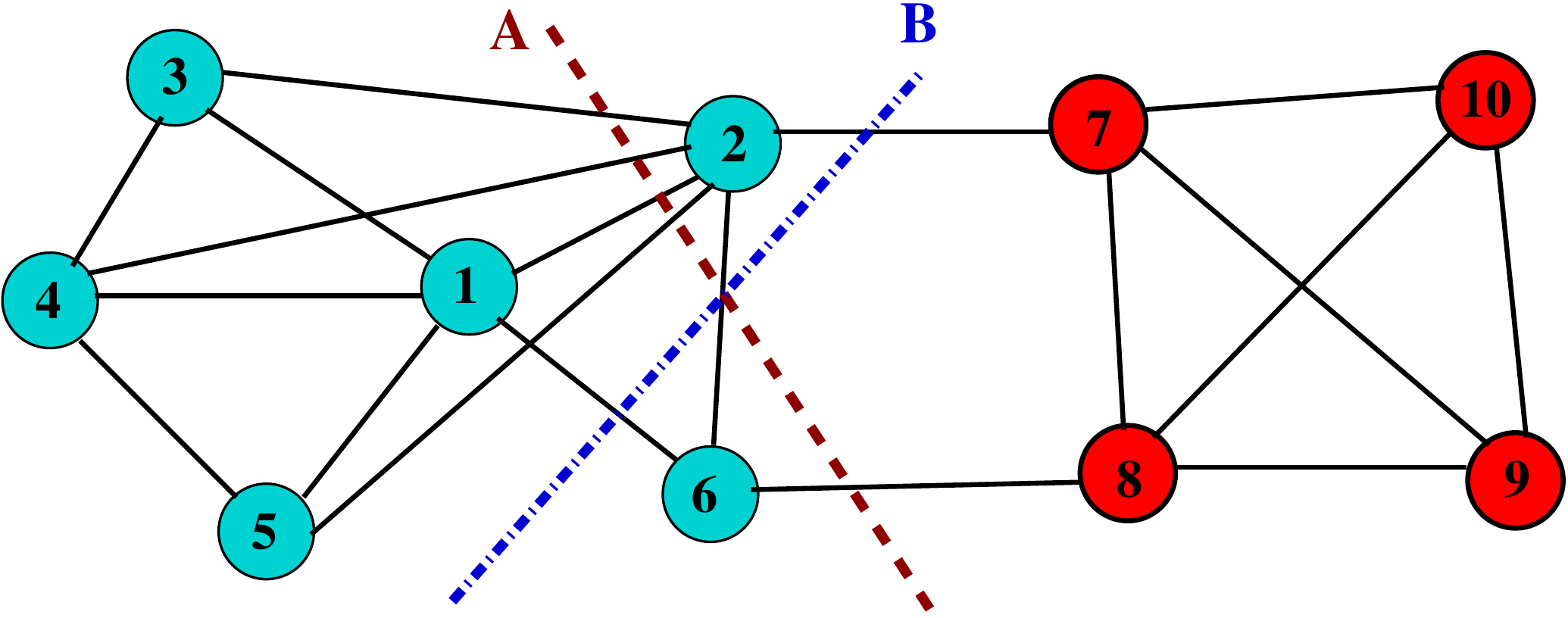}
 \caption{Example illustrating the limitation of purely partition-based measures. Colors correspond to the actual communities, whereas lines labeled and represent two different (incorrect) estimations of this community structure as presented in \cite{abs-1303-5441}.}
 \label{error}
\end{figure}

\cite{Orman12} argued that the traditional measures consider a community structure to be simply a partition, and  therefore ignore a part of the available information: the network topology. For instance, let us take an example in Figure \ref{error} where 10 nodes are partitioned into two communities (represented by two colors, and referred to as $R$). Let us also assume two detected community structures: ($A$) node 2 is misclassified and assigned with nodes, 7, 8, 9 and 10, ($B$) node 6 is misclassified and assigned with nodes, 7, 8, 9 and 10. If we apply the existing measures, in order to compare $R$ and $A$, we obtain the score for both the purity and inverse purity. Consequently, the F-Measure reaches the same value. For ARI and NMI, we get 0.6 and 0.62, respectively. Now, if we process the same measures between $R$ and $B$, we get exactly the same values. Indeed, from a partition perspective, nothing allows to distinguish node $2$ from node $6$, so misclassifying the former or the latter leads to the same evaluation. However, intuitively, these errors do not seem to be equivalent at all. Indeed, node $2$ is much more integrated in its actual community than node $6$. Its misclassification in partition $A$ is therefore a more serious error than that of node in partition $B$. 

\subsection{Issues with NMI}
A major problem of NMI is that it is not a {\em true metric}, i.e., it does not follow triangle-inequality. Imagine that a putative optimal partition is estimated according to a given criterion. Let us now consider the following triangle inequality: 
\begin{equation}\label{nmi_pro}
\frac{ NMI_{IE}+NMI_{EF}}{2} \leq \frac{1+NMI_{IF}}{2}
\end{equation}
where $NMI_{IE}$ is the normalized mutual information calculated for the initial structure ($I$) and the estimated partition ($E$), $NMI_{EF}$ is the normalized mutual information for the final structure ($F$) versus the estimated partition and $NMI_{IF}$ is the normalized mutual information for the comparison between the initial and final structures. Inequality \ref{nmi_pro} holds true if the structures of $I$, $F$ and $E$ are identical (i.e., both the number and sizes of the communities are the same, but not necessarily are the same the nodes within each community).

\begin{figure}%
\centering
\includegraphics[width=0.8\columnwidth]{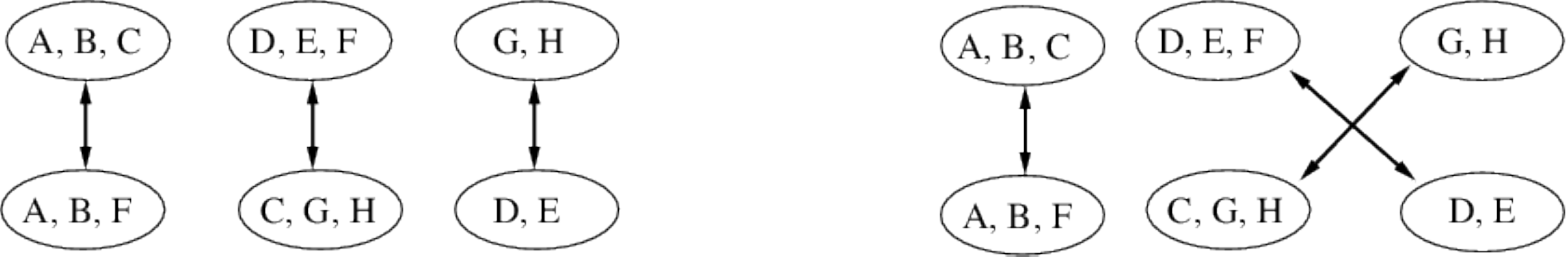}%
\caption{Example of the matching between parts of different partitions \cite{5520221}}%
\label{edit}%
\end{figure}

\subsection{Example of Edit Distance}
For example, Figue \ref{edit} presents two partitions with two possible matchings. In the first one, nodes $C$, $D$, $E$, $F$, $G$ and $H$ must be moved whereas, only $C$ and $F$ have to move with the second matching. The matching which minimizes the number of transformations can be computed in $O(n^3)$ with $n$ the number of parts using the Kuhn-Munkres algorithm \cite{Kuhn1955Hungarian}. 

\begin{figure}[t]
\centering
\begin{tabular}{@{}c@{}}
\includegraphics[width=\columnwidth]{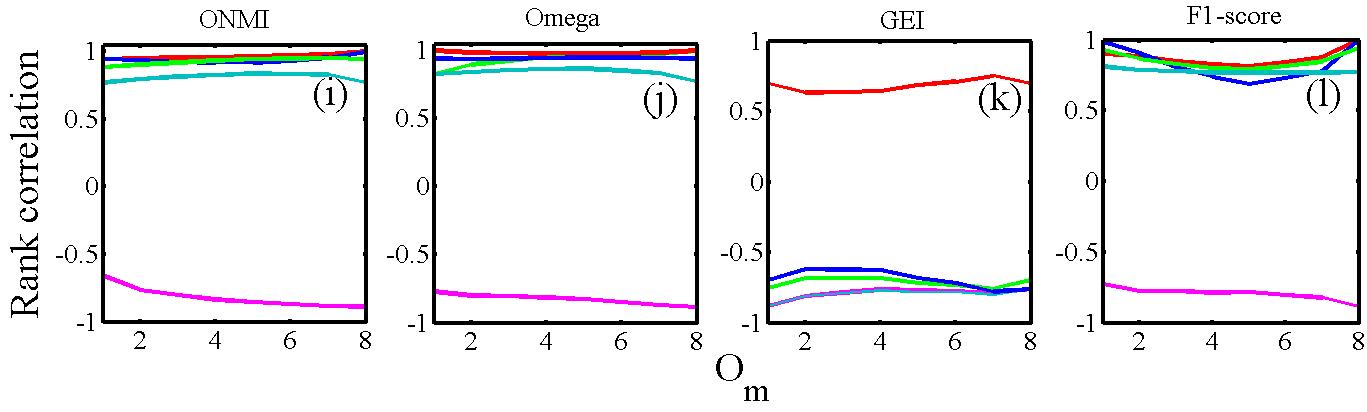}%
\end{tabular}
\centering
\begin{tabular}{@{}c@{}}
\includegraphics[width=\columnwidth]{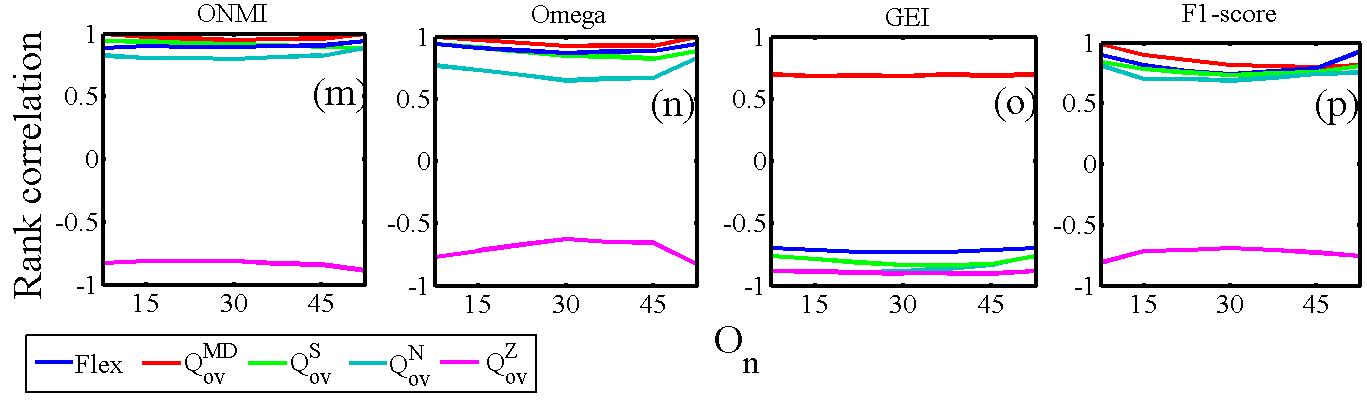}%
\end{tabular}
\caption{(Color online) Spearman's rank correlation among the results obtained from five overlapping community scoring metrics and four validation measures for LFR networks with overlapping community structure:  varying the number of communities to which a node belongs $O_m$ ($n=10,000$, $\mu=0.3$, $O_n=10\%$), and percentage of overlapping nodes $O_n$ ($n=10,000$, $\mu=0.3$, $O_m=5$).}
\label{lfr_over}
\end{figure}

\section{Experiments}
 We use the LFR benchmark networks proposed by \cite{Lancichinetti}. We vary the following parameters depending upon the experimental need: the number of nodes ($n$), mixing coefficient ($\mu$), the percentage of overlapping nodes $O_n$, and the number of communities to which a node belongs $O_m$. Figure \ref{lfr_over} shows the results for the LFR networks by varying different parameters $O_m$ and $O_n$.

\section{Summary}
In this section we summarize the various metrics for both community detection and evaluation. The notations used are borrowed from Table I in the main paper. Table I and Table II summarize the community detection metrics for non-overlapping and overlapping community structures respectively. Table III and Table IV describe the community evaluation metrics for non-overlapping and overlapping community structures respectively.

\end{document}